\documentclass[modern]{aastex631}
\usepackage{natbib}
\usepackage{amsmath}
\usepackage{graphicx}
\usepackage{diagbox}

\submitjournal{PASP}

\newcommand{\hi}{H\,\textsc{i}}
\newcommand{\cmfast}{\texttt{21cmFAST}}
\newcommand{\zreion}{\texttt{zreion}}

\newcommand{\edited}[1]{{\color{black}#1}}
\graphicspath{{./}{figures/}}

\begin{document}

\title{Understanding the Impact of Semi-numeric Reionization Models when using CNNs}

\author[0000-0002-8828-8461]{Yihao Zhou}
\affiliation{School of Science, Xi'an Jiaotong University, No. 28 West Xianning Road, Xi'an 710049, People's Republic of China}

\author[0000-0002-4693-0102]{Paul La Plante}
\affiliation{Astronomy Department, University of California, Berkeley, CA 94720, USA}
\affiliation{Berkeley Center for Cosmological Physics, University of California, Berkeley, CA 94720, USA}

\correspondingauthor{Yihao Zhou}
\email{yihaozhou20@gmail.com}

\begin{abstract}
Interpreting 21\,cm measurements from current and upcoming experiments like HERA and the SKA will provide new scientific insights and exciting implications for astrophysics and cosmology regarding the Epoch of Reionization (EoR). Several recent works have proposed using machine learning methods, such as convolutions neural networks (CNNs), to analyze images of reionization generated by these experiments since they could take full advantage of information contained in the image.
Generally, these studies have used only a single semi-numeric method to generate the input 21\,cm data. In this work, we investigate the extent to which training CNNs for reionization applications depends on the underlying semi-numeric models. Working in the context of predicting CMB $\tau$ from 21cm images, we compare networks trained on similar datasets from \cmfast\ and \zreion, two widely used semi-numeric reionization methods.
We show that neural networks trained on input data from only one model produce poor predictions on data from the other model. Satisfactory results are only achieved when both models are included in the training data. This finding has important implications for future analyses on observation data, and encourages the use of multiple models to produce images that capture the full complexity of the EoR.
\end{abstract}

\section{Introduction}
\label{sec:intro}

The Epoch of Reionization (EoR, $6 \lesssim z \lesssim 15$) refers to a period of the universe preceded by the Cosmic Dawn ($15 \lesssim z \lesssim 30$), when the first stars formed, and the Dark Ages ($30 \lesssim z \lesssim 200$), the period following recombination \citep{loeb_furlanetto2012}. During the EoR, some of the earliest luminous sources emitted a sufficient number of ultraviolet and soft X-ray photons that escaped into the intergalactic medium (IGM) such that it reionized neutral hydrogen \hi. As a result, the majority of baryonic matter in the IGM transitioned to a highly ionized state \citep{Furlanetto2016}, which is how it remains today. 

There are several ways to study this major phase transition: distant quasars, the cosmic microwave background (CMB), and the 21\,cm line of \hi. Analyzing the absorption properties of the IGM using quasars has provided important constraints on reionization, showing that reionization was likely only partially completed at $7 \lesssim z \lesssim 8$ \citep{schenker_etal2014,davies_etal2018,mason_etal2019}. Absorption spectra from lower redshift measurements argue that reionization was mostly completed by $z \sim 6$ \citep{Fan2006}, though other measurements may indicate that reionization was still ongoing as late as $z \sim 5.5$ \citep{bosman_etal2021}. Complementing this, the relatively low CMB optical depth $\tau$ \citep{planck2018} suggests that reionzation happened relatively late and was extended in redshift. Future detection of the redshifted 21\,cm line emission will be an essential probe of the thermal and ionization state of the IGM \citep{Fan2006,Furlanetto2016}. The neutral hydrogen gas in the IGM emits radiation due to the hyperfine transition of the \hi\ ground state, with a wavelength of $\lambda = 21$~cm in the hydrogen rest-frame \citep{Zaldarriaga_2004}. Currently, several experiments have been designed to measure the global temperature change of 21\,cm, such as the Experiment to Detect the Global EoR Signature (EDGES; \citealt{Bowman2018}), or target a statistical detection through the power spectrum, including the Low Frequency Array (LOFAR; \citealt{LOFRA}) and the Murchison Widefield Array (MWA; \citealt{MWA}). The Hydrogen Epoch of Reionization Array (HERA; \citealt{DeBoer_2017}) has recently provided the tightest constraints from an interferometer on the 21\,cm power spectrum \citep{hera2021}, which has ruled out part of the physically allowed phase space of galaxy models \citep{h1c_theory}. The Square Kilometre Array (SKA; \citealt{2015aska.confE...1K}) will provide high signal-to-noise imaging of the EoR and the Cosmic Dawn, mapping the first billion years of our universe.

The 21\,cm signal encodes a panoply of information about the first generation of galaxies, through which we can get constrains on key properties of reionization and gain an insight into the relating astrophysical process \citep{McQuinn2007,Fialkov2017}. However, how to extract information from the images generated by the arrays mentioned above still remains a great challenge
due to the strong foreground contamination that is several orders of magnitude larger than the EoR signal \citep{Morales2010,pober_etal2013}. 
Computing the power spectrum is \edited{a reliable way} given these measurement difficulties because the foreground contamination is intrinsically smooth while the target signal fluctuates rapidly. \edited{Several recent studies have developed robust methods for using the 21\,cm power spectrum to infer properties of early galaxies from the EoR \citep{Liu2016,Greig2018,Qin2020}. While because the 21\,cm field is highly non-Gaussian \citep{shimabukuro_etal2016,majumdar_etal2018}, in principle} there is more information that can be extracted by using a higher-point estimator such as the bispectrum, or by working in configuration space directly \edited{\citep{Hutter2020,watkinson_etal2022}}. By configuration space, we mean the real or redshift space data, which is in contrast to Fourier space.

One such solution to this problem is to use machine learning (ML) techniques. Machine learning is a powerful tool to find complex relationships among data without requiring much \textit{a priori} information about the functional forms of statistics used to extract the desired information. Crucially, because these methods work natively in configuration space, they are not limited to Gaussian information, allowing us in principle to take full advantage of all information in the 21\,cm maps. 
Machine learning has been used in a variety of works to analyze maps relevant to 21\,cm cosmology. For example, some studies have used ML to infer galaxy model properties from \cmfast\ \citep{Gillet2018}, model-independent parameters such as the duration of reionization $\Delta z$ \citep{LaPlante2018}, or cosmological parameters such as $\tau$ and $\sigma_8$ \citep{hortua_etal2020,billings2021}. 
Recently, \citet{Zhao2022_DELFI} use 3D CNN to compress 3D image data and perform a Bayesian inference of reionization parameters, which outperforms earlier analysis based on two-dimensional 21cm images.
Other studies have attempted to reconstruct the underlying ionization field \citep{gagnon-hartman2021} or dark matter density field \citep{villanuevadomingo2021} from single-model input data. 
Besides that, generative adversarial networks (GANs) are used in some works to generate maps of neutral hydrogen \citep{Zamudio-Fernandez2019} or predicted different summary statistics using CNN architectures \citep{Kwon2020}.

In most studies mentioned above, a semi-numeric method is used to convert the initial conditions and dark matter density fields into ionization fields and ultimately 21\,cm maps. Two particular frameworks to model and simulate reionization are \cmfast\ \citep{Mesinger2011,21cmfast_v3} and \zreion\ \citep{Battaglia2013}, which produce qualitatively similar yet distinct predictions. Generally, only one semi-numeric method is used to generate testing and training data without a detailed treatment of the uncertainty induced by using solely one method.
In order to understand the applicability of ML techniques to real observation data in the future, we must endeavor to build and train networks that are as flexible and robust as possible. In particular, these networks should be insensitive to the particular details of a specific semi-numeric method when predicting a model-independent parameter such as $\tau$ or $\sigma_8$. In other words, the network is supposed to be independent of specific semi-numeric EoR models. We aims to investigate this problem in this work in the context of regressing optical depth to the CMB optical depth. We compare the behavior of convolutional neural networks (CNNs) on \zreion\ and \cmfast. We also discuss the possibility of having one network that works on both of the two models. 

We organize this paper as follows. In Section~\ref{sec:simulations}, we introduce and compare two semi-numeric models used in this work. In  Section~\ref{sec:Method}, we describe the ML techniques we employed. In Section~\ref{sec:Results}, we demonstrate the performances of the CNN on the two models. We discuss the results in Section~\ref{sec:discussion} and conclude in Section~\ref{sec:conclusion} with a summary and avenues for future study. Throughout the work, unless stated otherwise, we employ a $\Lambda$CDM cosmology with cosmological constants consistent with the Planck 2018 results \citep{planck2018}.

\section{Semi-numeric Reionization Methods}
\label{sec:simulations}

Accurately simulating the EoR with sufficient fidelity requires a delicate balancing act between enormously different length scales and the sophistication of the relevant physics. On small-scales, the formation of Population~III and Population~II stars as the predominant engines driving reionization happens on the parsec (pc) scale \citep{Regan2020}. Alternatively, the typical size of ionized bubbles during the EoR is tens to hundreds of Mpc. This severe difference in length scale means that accurately resolving all relevant physics is generally not feasible given current computational resources. Furthermore, the complicated nature of reionization generally requires to include $N$-body methods and hydrodynamics for galaxy formation, and radiative transfer to capture the large-scale structure, the feedback between photons and baryons in the IGM. 

Considering that upcoming experiments are expected to have relatively poor angular resolution, large-scale predictions that are accurate on Mpc scales are generally sufficient for understanding and interpreting 21\,cm measurements. 
Because it is most important to capture the largest scales, various approximations can be made.
In particular, in the literature there are several semi-numeric simulation methods that do not require full radiative transfer simulations to generate predictions for reionization that are reasonably accurate. Two specific ones are \zreion\, described in \citet{Battaglia2013}, and \cmfast\ described in \citet{Mesinger2011}. Below, we briefly summarize their key features and refer the reader to the original publications for a more detailed discussion of each model.

\subsection{\zreion}
\label{sec:zreion}

The \zreion\ method\footnote{\url{https://github.com/plaplant/zreion}} is presented in \citet{Battaglia2013}, and has been applied to the kinetic Sunyaev-Zel'dovich effect \citep{Battaglia2013,natarajan_etal2013} and the 21\,cm field \citep{LaPlante2014,LaPlante2018,LaPlante2020}. The method treats the fluctuation of ``redshift of reionization field'' $\delta_{z}(\boldsymbol{x})$ as a biased version of the matter overdensity field $\delta_{m}(\boldsymbol{x})$ on large scales ($\gtrsim 1\ h^{-1}\mathrm{Mpc}$). First, we define $\delta_{m}(\boldsymbol{x})$ as 
\begin{equation}
\delta_{m}(\boldsymbol{x}) \equiv \frac{\rho_{m}(\boldsymbol{x})-\bar{\rho}_{m}}{\bar{\rho}_{m}},
\end{equation}
where $\bar{\rho}_{m}$ is the mean matter density, and $\delta_{z}(\boldsymbol{x})$ as  
\begin{equation}
\delta_{z}(\boldsymbol{x}) \equiv \frac{\left[z_{\mathrm{re}}(\boldsymbol{x})+1\right]-[\bar{z}+1]}{\bar{z}+1},
\end{equation}
where $\bar{z}$ is the mean value for the $z_{\mathrm{re}}(\boldsymbol{x})$ field. 
\zreion\ takes the advantage of the fact that $\delta_{m}(\boldsymbol{x})$ is highly correlated with $\delta_{z}(\boldsymbol{x})$. We introduce a bias parameter $b_{zm}(k)$ to quantify the relationship between the two fields in Fourier space:
\begin{equation}
b_{z m}^{2}(k) \equiv \frac{\left\langle\delta_{z}^{*} \delta_{z}\right\rangle_{k}}{\left\langle\delta_{m}^{*} \delta_{m}\right\rangle_{k}}=\frac{P_{z z}(k)}{P_{m m}(k)},
\end{equation}
where $P_{xx}(k)$ is the power spectrum of $\delta_{x}$ and $P_{zz}(k)$ is that of $\delta_{z}$.
To parameterize the bias, the method uses two parameters $k_0$ and $\alpha$ to write $b_{z m}$ as a function of Fourier wavenumber $k$:
\begin{equation}
b_{z m}(k) = \frac{b_{0}}{\left(1+\frac{k}{k_{0}}\right)^{\alpha}}.
\end{equation}
We use the value of $b_{0}=1 / \delta_{c}=0.593$, where $\delta_c$ is the critical overdensity in the spherical collapse model. Given a cosmological density field, the method relies only on the value of the three parameters $\{\bar{z}, k_0, \alpha\}$ to determine $z_\mathrm{re}(\boldsymbol{x})$. The midpoint of reionization depends largely on the mean value $\bar{z}$ (though they are not equivalent in general unless the mean and median redshift are the same), and the duration is controlled by $k_0$ and $\alpha$.

In the original work of \citet{Battaglia2013}, the values of $\bar{z}$, $k_0$, and $\alpha$ were calibrated by hydrodynamic simulations of reionization with radiative transfer. Using this approach, the best-fit values for $k_0$ and $\alpha$ were $k_0 = 0.185$ $h\mathrm{Mpc}^{-1}$ and $\alpha = 0.564$. However, the method is sufficiently flexible that varied values could be used when generating the reionization field $\delta_\mathrm{z}(\boldsymbol{x})$. This flexibility allows for a broad range of different reionization scenarios to be generated relatively quickly. Nevertheless, there are some limitations associated with this method. Due to the fact that this method does not explicitly use the sources of ionizing radiation when generating the reionization history, it is not straightforward to include specific models of galaxy formation. Instead, reionization is treated in a statistical manner, without considering the properties of individual galaxies. Also, due to the treatment of the ``redshift of reionization'' field as a biased tracer of the dark matter field, it is difficult to generate reionization histories that are very extended in duration, or very asymmetric about the midpoint of reionization.  For instance, \zreion\ could hardly produce a reionization history that has an early low-level of ionization ($\sim 10\%$) for a long time before the midpoint of reionization, as might be caused by a significant amount of radiation from Pop~III stars \citep{miranda_etal2017}. It also does not easily account for quasar-like sources that generate large ($\sim 100\ h^{-1}\mathrm{Mpc}$) ionized regions. This is largely due to the fact that the dark matter field is only weakly non-Gaussian on large scales at high redshift, and the resulting $\delta_{z}(x)$ will also be only weakly non-Gaussian.

Despite these limitations, \zreion\ can produce similar reionization histories to those from \cmfast\ or hydrodynamic simulations with much less time.
Another benefit of \zreion\ as a method is that it treats the dark matter density field as an external input, which facilities an apples-to-apples comparison with other methods like \cmfast. It also supports running on matter fields generated from $N$-body simulations, as was originally done in \citet{Battaglia2013}, or from second-order Lagrangian perturbation theory (2LPT). 

\subsection{\cmfast}
\label{sec:21cmfast}

The \cmfast\ method\footnote{\url{https://github.com/21cmfast/21cmFAST}} \citep{Mesinger2007,Mesinger2011,21cmfast_v3} assumes galaxy-driven reionization and includes a more involved treatment of reionization sources. Using the excursion set formalism \citep{Bond1991} and Lagrangian perturbation theory, the dark matter distribution of a volume can be converted into a ``halo field'' that form the sources of ionizing radiation. Starting from initial conditions, the halo field is computed in a comoving volume at decreasing redshift values. Given the halo field at a particular redshift, \cmfast\ uses a series of astrophysical parameters pertaining to galaxy and star formation to compute the corresponding radiation field. Once the cumulative number of hydrogen-ionizing photons has been calculated from UV and X-ray sources, the number of neutral atoms surrounding these sources are computed within regions of decreasing $R$, which begins at a parameter defining the maximum mean-free path of ionizing radiation, $R_\mathrm{mfp}$, and decreases down to the resolution of an individual cell, $R_\mathrm{cell}$. A voxel is considered fully ionized when \citep{Greig2018}
\begin{equation}
\zeta f_\mathrm{coll}(\mathbf{x},z,R,\bar{M}_\mathrm{min}) \geq 1,
\label{eqn:fcoll}
\end{equation}
where $f_\mathrm{coll}(\mathbf{x},z,R,\bar{M}_\mathrm{min})$ is the fraction of collapsed matter residing within halos more massive than $M_\mathrm{min}$ \citep{Press1974,Bond1991,Lacey1993,Sheth1999}, and $\zeta$ is an ionizing efficiency factor which accounts for how dark matter halos convert gas into ionizing photons.

This approach assumes a constant relationship between the ionizing luminosity and the halo mass down to some minimum mass $\bar{M}_\mathrm{min}$, below which star formation is inefficient due to feedback and/or cooling effects. This mass can be expressed in terms of a redshift-dependent halo virial temperature $T_\mathrm{vir}^\mathrm{min}$ \citep{Barkana2001}:
\begin{multline}
M_\mathrm{vir}^\mathrm{min}(z) = 10^8 h^{-1} \left(\frac{\mu}{0.6}\right)^{-3/2} \left(\frac{\Omega_{\mathrm{M},0}}{\Omega_\mathrm{M}(z)} \frac{\Delta_c}{18\pi^2} \right)^{-1/2} \\
\times \left(\frac{T_\mathrm{vir}^\mathrm{min}}{1.98 \times 10^4 \mathrm{K}} \right)^{3/2} \left(\frac{1+z}{10} \right)^{-3/2} \mathrm{M}_\odot,
\end{multline}
where $\mu$ is the mean molecular weight, $\Omega_\mathrm{M}(z) = \Omega_{\mathrm{M},0} (1+z)^3/\left[ \Omega_{\mathrm{M},0} (1+z)^3 + \Omega_\Lambda \right]$, and $\Delta_c = 18\pi^2 + 82d - 39d^2$, and $d = \Omega_\mathrm{M}(z) - 1$.


Once the the relations between halo mass and ionizing efficiency are defined, $k$-space filtering is done to determine the scales at which the amount of ionization radiation produced is equal to the amount of neutral hydrogen gas, which informs which regions have been ionized. Following this, the ionization field is generated and is then converted into a 21\,cm brightness field. For our purposes, we extract the ionization history given this ionization field as well as snapshots of the 21\,cm brightness temperature. The midpoint and duration of reionization are thus not controlled directly by the user, and instead are the result of the complicated interplay of the different astrophysical parameters.

\subsection{Comparison}
\label{sec:comparison}
In the following analysis, we generate snapshots of the 21\,cm field using both \zreion\ and \cmfast\ that serve as training data input for machine learning methods (described more below in Sec.~\ref{sec:Method}). We use the same cosmology, comoving box size of 2 $h^{-1}$Gpc, and 512 output resolution elements per linear dimension for both simulations. The input data are two-dimensional images of the 21\,cm brightness temperature snapshots generated at 30 different redshift slices, which are the same for the two methods. 

\begin{figure}[t]
  \centering
  \includegraphics[width=0.95\textwidth]{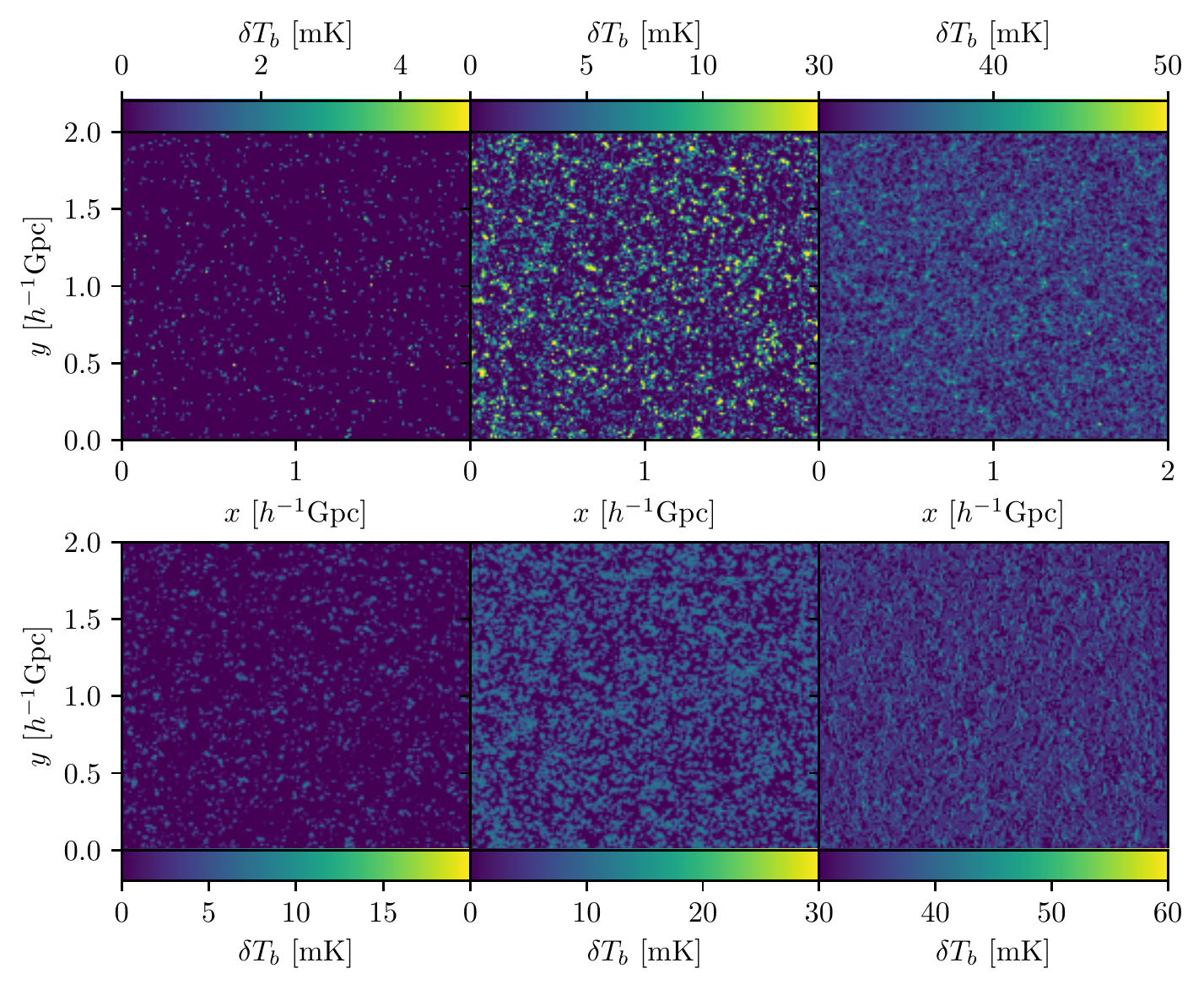}
  \caption{A comparison of \zreion\ (top) and \cmfast\ (bottom) simulations. The columns show a two-dimensional snapshot at $z=6$ (left), $z=7$ (center), and $z=15$ (right). Because the same density field is used for the two simulations, the differences are due primarily to the means by which the ionization and brightness temperature fields are computed. Note that although the small scales of the images are different, especially at relatively low redshift values, the larger scale features are generally comparable.}
  \label{fig:t21_vis}
\end{figure}

There are two primary differences between \zreion\ and \cmfast\ methods: the means of generating the underlying dark matter density field ($N$-body simulations for \zreion, second-order Lagrangian perturbation theory (2LPT) for \cmfast) and the treatment of ionization sources (implicit sources for \zreion, explicit galaxies for \cmfast). As mentioned  in Sec.~\ref{sec:21cmfast}, \cmfast\ uses 2LPT to generate a comparatively low-resolution grid from the high-resolution Gaussian initial conditions at a very high redshift. The ionization field and associated calculations are implemented on this lower-resolution field. For this study, \zreion\ uses $N$-body simulation snapshots and work all the way on the high-resolution field until the last step before outputting. which makes \zreion\ to have more accurate density fields at the cost of computation time. For the method to generate brightness temperature, \cmfast\ works mostly in configuration space while \zreion\ works in Fourier space. 

\begin{figure}[t]
\centering
\includegraphics[width=0.7\textwidth]{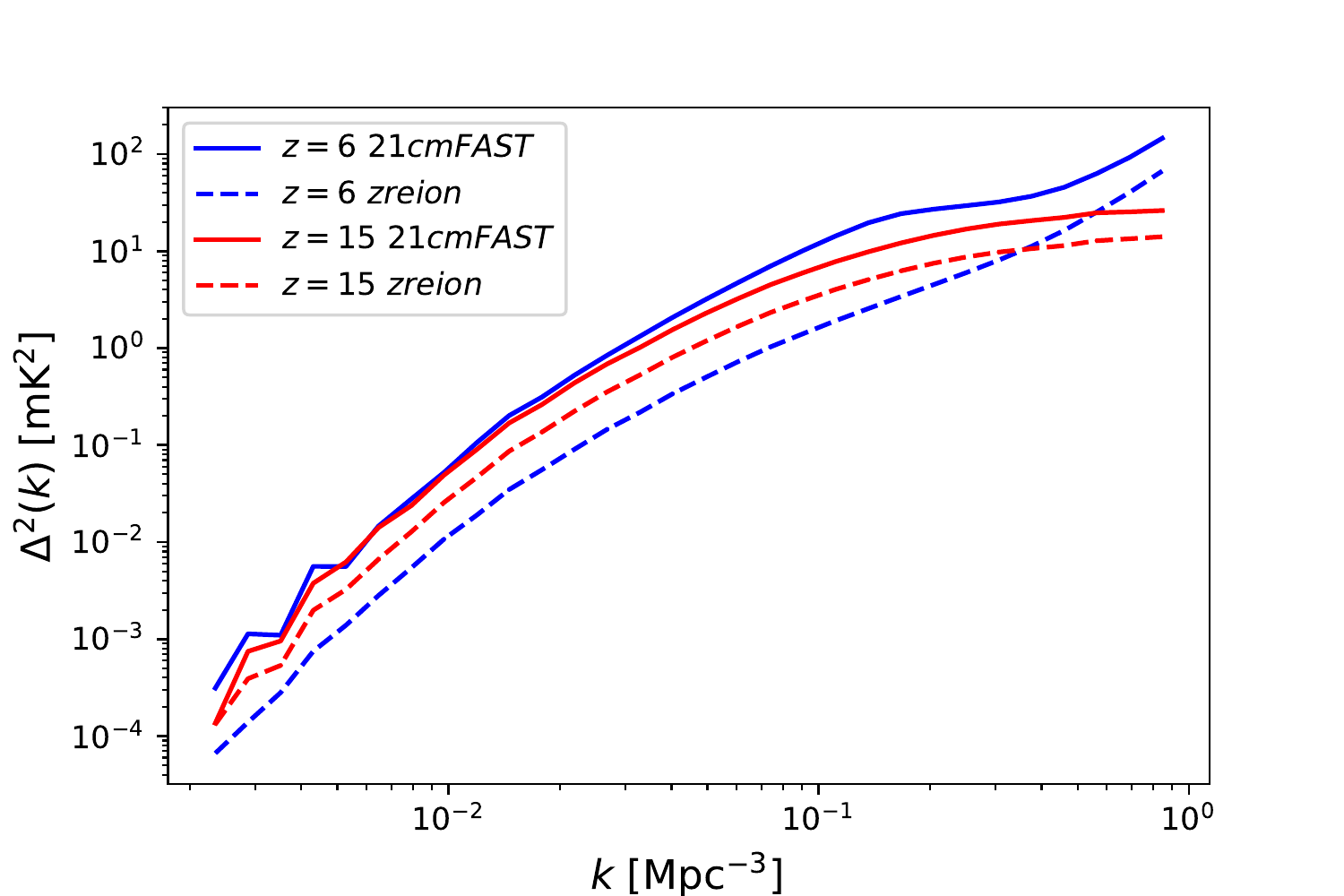}
\caption{The power spectra} of the 21\,cm brightness temperature for \zreion\ and \cmfast\ simulations at low redshift ($z=6$) and high redshift ($z=15$). They are generated using the same density fields. In addition to the slightly different overall amplitudes, there are also distinct shapes at small scales (large $k$ values).
\label{fig:power_spectrum}
\end{figure}

Figure~\ref{fig:t21_vis} shows a qualitative comparison of the 21\,cm brightness temperature as generated by \zreion\ and \cmfast\ using the same underlying density field and a comparable reionization history. As can be seen, the two simulations are capable of producing comparable results since the general topology of the ionization field is similar between the two methods except the higher peaks in the \zreion\ simulation at lower redshift.
The 21\,cm power spectra at high redshift ($z=15$) and low redshift ($z=6$) shown in Figure~\ref{fig:power_spectrum} provide another comparison of the \zreion\ and \cmfast\ in Fourier space.
At high redshift, the shapes of the power spectra are very similar, though they have an overall normalization difference. At low redshift, however, there is a difference both in amplitude and the shape as a function of $k$. 
Without considering the general amplitude level, \cmfast\ has relatively less power for large $k$ values (smalls scales) compared with \zreion.
Since different radiation feedback scenarios, i.e., varied parameter sets, can hardly affect the power spectra at high $k$ value, especially near the end of reionization, \citep{Hutter2021_Astraeus},
the distinctions in configuration space and Fourier space could both be attributed to the different methods of generating the ionization fields and the brightness temperature, particularly with respect to the low-density regions that are partially or totally neutral near the end of reionization.

Although the large-scale differences can be made to agree through an overall renormalization, the divergent behavior on small scales has
important impacts on generalizability when making predictions on parameters that are ostensibly model-independent, which we show more below in Sec.~\ref{sec:Results}.

\subsection{Generating Snapshots} 
\label{section:snapshots}
As mentioned previously, our input data 
are a series of two-dimensional snapshots of the 21\,cm brightness temperature generated at different redshift values. Crucially, without directly using the brightness temperature generated by \zreion\ or \cmfast, we apply two instrumental artifacts: a maximum angular resolution and a ``wedge filter'', which will be present in upcoming data generated by the HERA or the SKA. For the present study, we adopt the design parameters of HERA in this work.

The angular resolution of radio arrays is determined by the longest baselines: 
\begin{equation}
\mathbf{k}_{\perp,\mathrm{max}
}=\frac{2 \pi \mathbf{u}_{\mathrm{max}}}{D_{c}},
\end{equation}
where $D_{c}$ is the comoving distance along the line of sight and  $\mathbf{u}_{\mathrm{max}}$ is the maximum baseline vector. The longest baseline for HERA is $\sim870$m \citep{DeBoer_2017}, which corresponds to $ k_{\perp,\mathrm{max}} \sim 0.5\ h^{-1}\mathrm{Mpc}$ at $z = 8$. To include this effect in our snapshot data, we set all the modes with $k_{\perp} \geq k_{\perp,\mathrm{max}}$ to be 0 in Fourier space.

\begin{deluxetable*}{cccccc}[t]
\tablenum{1}\label{table:21cmfast_astro_para}
\tablecaption{Varied Astrophysical Parameters Used to Generate \cmfast\ and \zreion\ Dataset}
\tablewidth{0pt}
\tablehead{
\colhead{Parameter} & \colhead{Unit} &\colhead{Fiducial Value} & \colhead{$\sigma$} & \colhead{Minimum Value} & \colhead{Maximum Value}}
\startdata
$\log(T_{\mathrm{vir}}^{\mathrm{min}})$  & K   & 5.10 & 0.15 & --- & ---  \\
$\zeta$  & ---  & 100      & 25  & --- & ---  \\
$\log(\frac{M_{\mathrm{turn}}}{M_{\odot}})$    &  ---   & 8.7     & 0.3 & --- & --- \\
$E_{0}$   & keV  & 500     & 50  & --- & --- \\
$\alpha_{\mathrm{X}}$  & ---  & 1.0  & 0.3 & --- & --- \\
$t_{\star}$  & ---   & 0.5     & 0.1 & --- & --- \\
$\log(\frac{L_\mathrm{X}}{\mathrm{SFR}})$   & $\mathrm{erg}\ \mathrm{s}^{-1}\ \mathrm{M}_{\odot}^{-1}\ \mathrm{yr}$  & 40  &  0.5 & --- & --- \\ \hline
$\bar{z}$ & --- & --- & --- & 7 & 9 \\
$\alpha$ & --- & --- & --- & 0.05 & 0.2 \\
$k_0$ & $h$Mpc$^{-1}$ & --- & --- & 0.5 & 1.5 \\
\enddata
\tablecomments{For the \cmfast\ snapshots, we randomly choose a value from a Gaussian distribution for each parameter characterized by the mean (``fiducial'') value and $\sigma$ value specified in the table. For the \zreion\ snapshots, we randomly choose each parameter from a uniform distribution between the ranges of values. Choosing the model parameters in this way leads to ionization histories that are comparable, which leads to comparable values of $\tau$.}
\end{deluxetable*}

Another observation effect is foreground contamination, which is present in all interferometer data. Within the target wavelength range, the foreground radio signal, which mainly comes from the synchrotron radiation in the Milky Way, overwhelms the EoR signal by several orders of magnitudes \citep{Morales2010}. 
Since the galactic synchrotron radiation roughly follows a power law, it is intrinsically smooth in Fourier space. Na{\"\i}vely, the contamination should be limited to small $k_{\parallel}$ modes. However, due to the interferometer's chromatic response, the foreground signal scatters into high $k_{\parallel}$ modes and creates a ``wedge'' in the Fourier space \citep{Parsons2012}. The slope of the wedge $m(z)$ is a function of the baseline length and cosmology, and can be expressed as \citep{Thyagarajan2015}:
\begin{equation}
m(z) \equiv \frac{k_{\|}}{k_{\perp}}=\frac{\lambda D_{c} f_{21} H(z)}{c^{2}(1+z)^{2}},
\label{eqn:wedge_slope}
\end{equation}
where $\lambda$ is the 21\,cm signal wavelength at the given redshift $z$, $D_{c}$ is the comoving distance, $f_{21}$ is the rest-frame 21\,cm signal frequency, and $H$ is the Hubble constant. Note that the expression in Equation~(\ref{eqn:wedge_slope}) accounts for maximal data contamination; physically, the bright foreground contamination extends down to the horizon of the interferometer beam. To include this effect, we transform the snapshots into Fourier space and set all the Fourier modes where $k_{\parallel} \leq m(z) k_{\perp}$ to be 0. For the redshift values of interest here, $m \sim 3$, which leads to removing a significant amount of data from the input snapshots.

The way we generate snapshots is modeled after \citet{LaPlante2018}. For both \zreion\ and \cmfast, we vary model parameters given in Table~\ref{table:21cmfast_astro_para} to generate new simulations. For \cmfast, we use a Gaussian distribution for each parameter with the mean value listed in the third column and the standard deviation present in the fourth column. For \zreion, we randomly change $\{\bar{z}; k_0; \alpha\}$ with the minimum and maximum values shown in the fifth and sixth columns. Then, we generate brightness temperature boxes at a given redshift. After this we apply filters to include the observation effects mentioned above. We pick the central two-dimension slab in the box to represent the signal at a specific redshift. One slab lies in the plane of the sky, spanning $2 \ h^{-1} \mathrm{Gpc}$ with 512 pixels on each side. The axis along the line of sight spans $\sim 48.8\ h^{-1} \mathrm{Mpc}$. Over such relatively short distance the redshift evolution could be ignored, so the slab could be considered comoving and we do not include light cone effect \citep{LaPlante2014}. We generate 30 slices that extend over the redshift range $6 \le z \le 15$ and use them as a single image fed into the CNNs. Finally, we normalized each snapshots among $0 \sim 1$. As a result, each image has 512 $\times$ 512 $\times$ 30 pixels. We generate 1000 snapshots for \zreion\ and \cmfast, respectively. For both of the two methods, we keep the underlying cosmological parameters fixed, though we explore the impact of changing these parameters in Appendix~\ref{section:cosmology}.

\begin{figure}[t]
\centering
\includegraphics[width=0.95\textwidth]{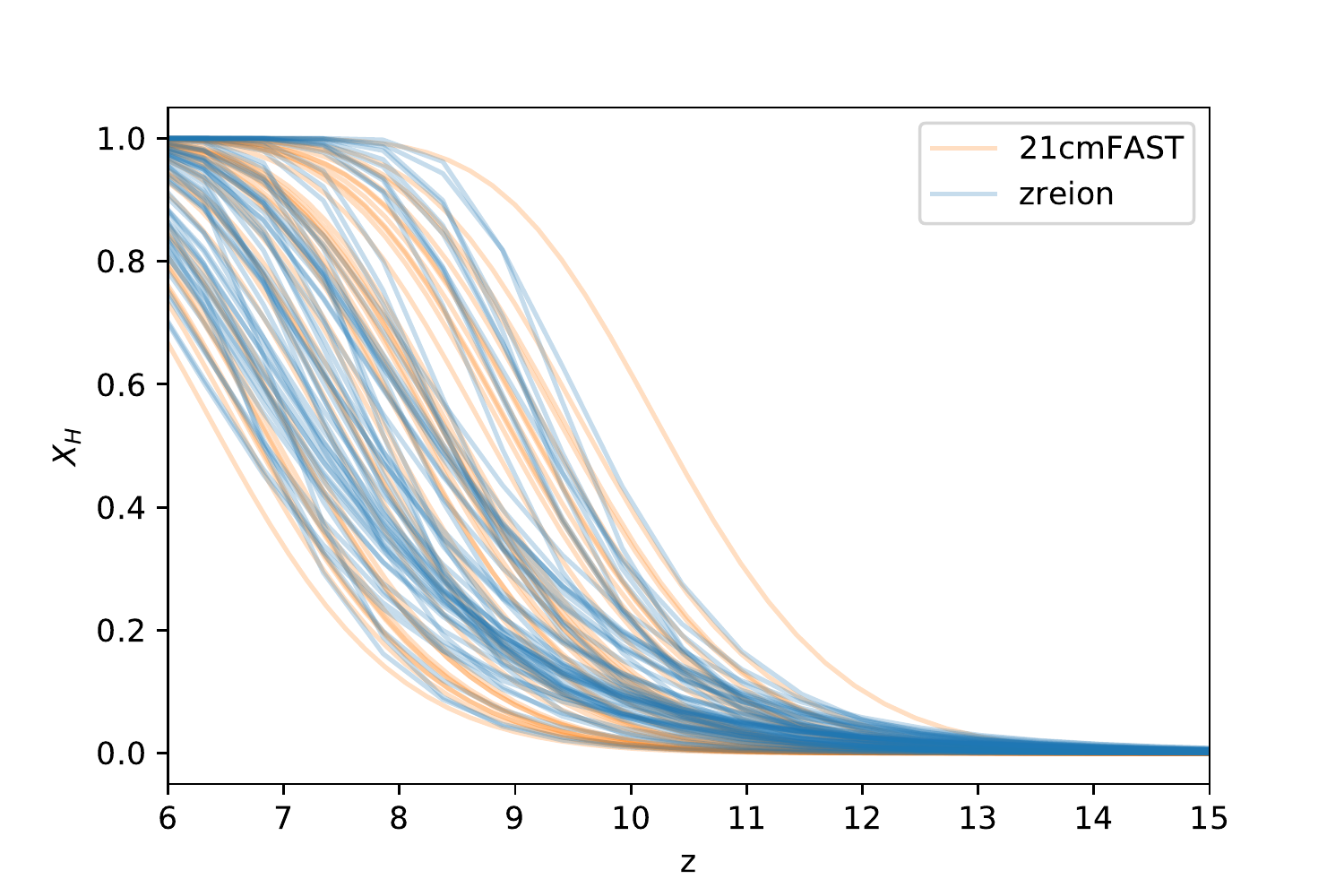}
\caption{The reionization history for some \zreion\ and \cmfast\ samples randomly chosen from the dataset. We show the ionization fraction $x_H$ as a function of redshift $z$. Qualitatively, the two methods produce similar histories.}
\label{fig:reionization_history}
\end{figure}

Figure~\ref{fig:reionization_history} shows the reionization history for several snapshots from \zreion\ and \cmfast, which are randomly chosen from the dataset. As can be seen, the histories produced by the two methods are qualitatively similar, and the large variances of parameter space produce a wide range of reionization histories. Despite the large-scale agreement between the two simulations, there are nevertheless several differences. In general, the reionization histories of \zreion\ are slightly briefer than \cmfast\ (e.g., the slope of the blue lines in the figure are mostly steeper than the orange ones). As discussed above in Sec.~\ref{sec:zreion}, it is difficult to generate \zreion\ simulations with very extended reionization histories. As such, there are several \cmfast\ samples with large duration that cannot be matched despite widely varying the \zreion\ parameters. Nevertheless, at the level of these global one-point statistics, the two methods can produce similar reionization histories, as well as the two-point level shown in Figures~\ref{fig:t21_vis} and \ref{fig:power_spectrum}.

\begin{figure}[t]
\centering
\includegraphics[width=0.95\textwidth]{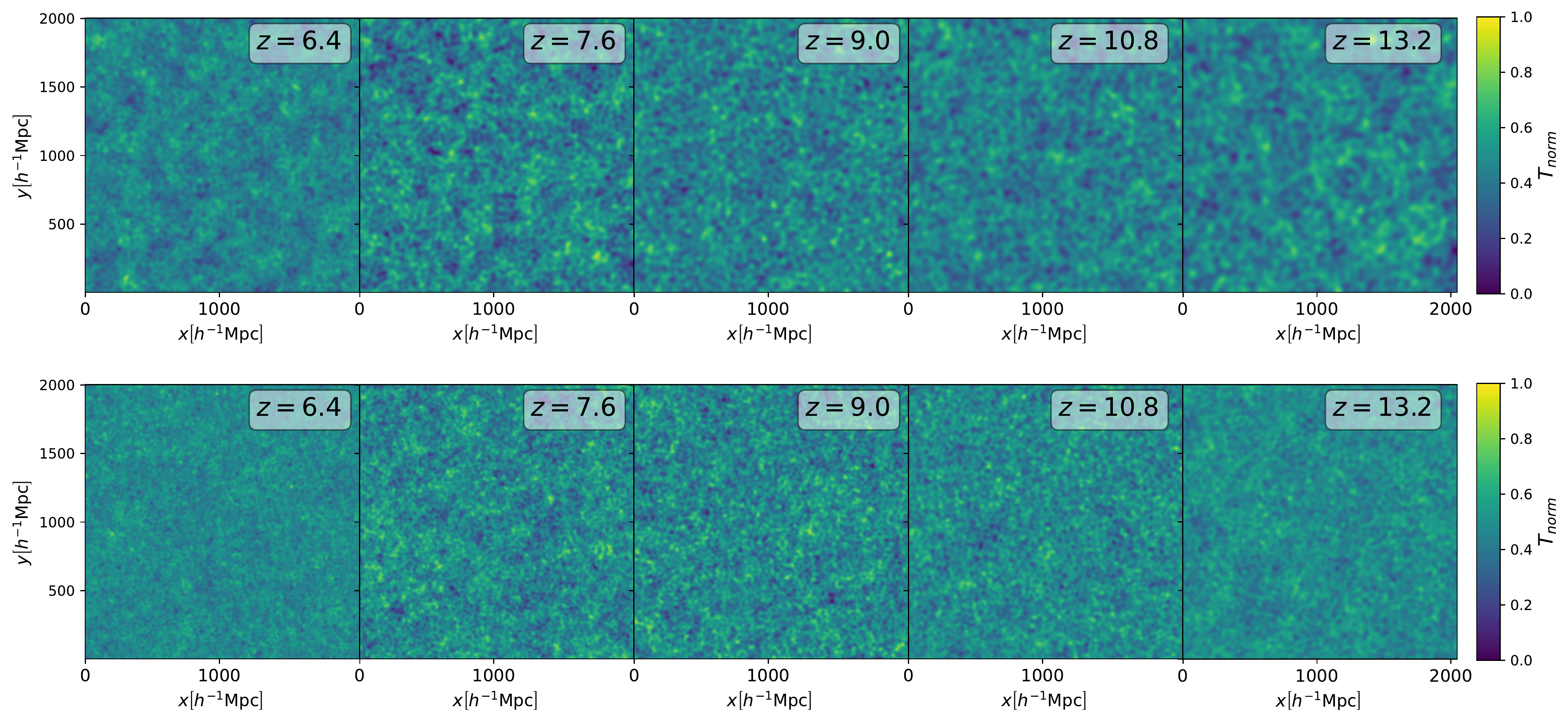}
\caption{A visualization of the \edited{normalized} 21\,cm brightness temperature snapshots for \cmfast\ (top) and \zreion\ (bottom) after the wedge filter has been applied. 
The different columns \edited{correspond to} different redshift slices.}
\label{fig:snapshots}
\end{figure}

In our case, the redshift axis serves as the color channels of the CNNs inputs. This is in contrast to other works, such as \citet{Gillet2018}, where the authors use two-dimensional input with one axis representing evolution along the line-of-sight (LOS). The latter approach may be easier for CNNs to capture the map-space features along the redshift/LOS axis and regress on global parameters such as reionization history and optical depth, because the redshift evolution is included explicitly as one of the axes of the image. However, actual observation data are difficult to reshape into such a two-dimensional format. Furthermore, CNNs conventionally deal with square-shaped images. For these reasons, we choose to use color channels to represent the redshift axis.
Figure~\ref{fig:snapshots} shows a comparison \edited{between a typical input \zreion\ sample and a \cmfast\ sample, which share similar reionization history and optical depth.} In contrast to Figure~\ref{fig:t21_vis}, these snapshots have had the wedge filter applied. Since we remove the $k=0$ mode to preclude referencing an absolute scale, Figure~\ref{fig:snapshots} shows \edited{variation} about the mean temperature of the map. 
Visually, the simulations outputs of \cmfast\ and \zreion\ are qualitatively \edited{comparable}. However, there are still important distinctions, such as the snapshots from \cmfast\ having larger fluctuation scale. 

\begin{figure*}
\centering
\includegraphics[width=0.48\textwidth]{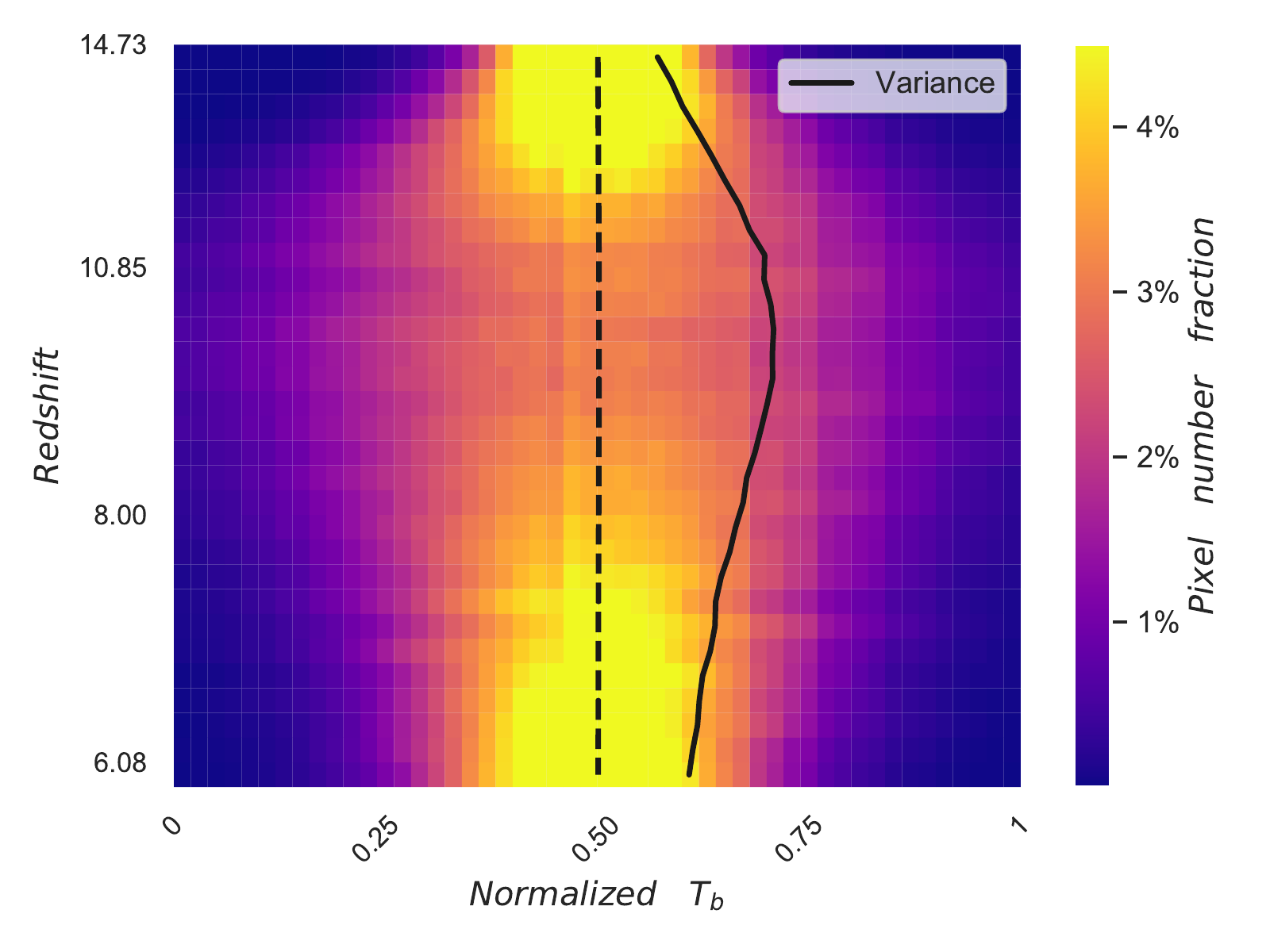}\hfill
\includegraphics[width=0.48\textwidth]{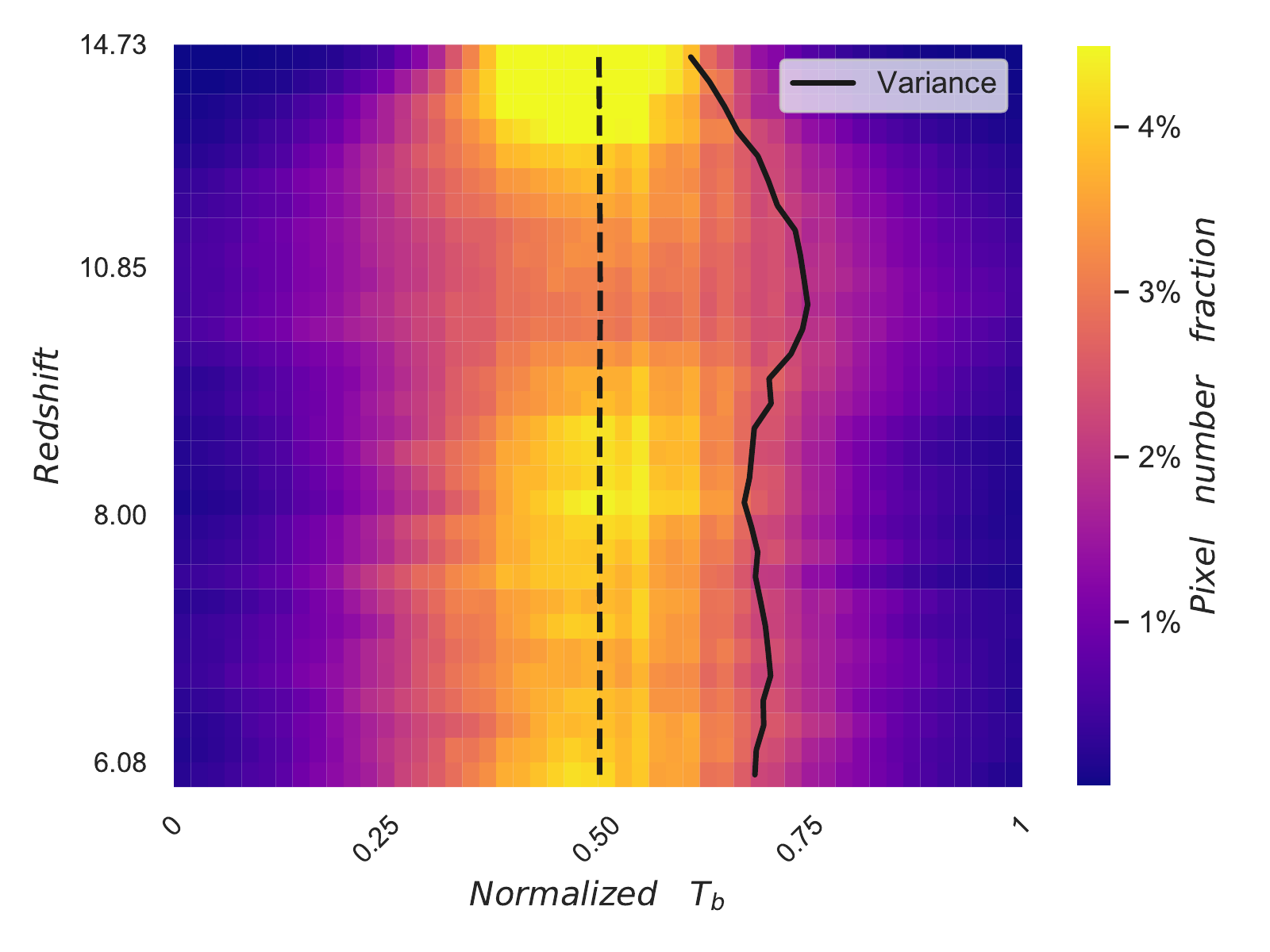}
\caption{The pixel distribution of \edited{one typical \zreion\ snapshot (left) and one \cmfast\ snapshot (right) used in the respective training sets}. These \edited{plots} show the fluctuation of the brightness temperature $\delta T_b$ normalized to be between 0 and 1. The $x$-axis corresponds to the temperature value and the $y$-axis corresponds to the redshift value. The solid black lines \edited{are} the variance as a function of redshift. The dash black lines \edited{label} the mean. See Sec.~\ref{section:snapshots} for further discussion.}
\label{fig:pixel_distribution}
\end{figure*}

Figure~\ref{fig:pixel_distribution} shows \edited{the normalized pixel values distribution for each slice in a \zreion\ snapshot and a \cmfast\ snapshot, which is a more quantitative comparison between the two methods}.
As mentioned above, there are 30 two-dimensional slices in each snapshot representing 30 different redshift values. \edited{Since we have} remove the $k_\perp=0$ mode, what we show here is essentially the distribution of fluctuations, \edited{which are encoded by the color.} 
The solid black line \edited{indicates} the variance in brightness temperature for a given redshift value, with the dashed line serving as a reference for the mean value.
At the beginning of reionization, most of the simulated space is unionized so pixels are concentrated around the mean value. When reionization is partially underway, the variance increases because the ionization state of the volume starts to show large variations. Near the end of reionization, space becomes homogeneous again because most of the volume is ionized. As can be seen, \edited{the snapshots produced by \zreion\ has} a variance that is larger near the midpoint of reionization, and is smaller at the end. Conversely, for \cmfast\ snapshots the pixel variance stays relatively large at the low redshift. 

\section{Method}
\label{sec:Method}
Convolutional neural networks (CNNs; \citealt{Fuku1982,Lecun1999_CNN,Krizh_CNN}) are a family of machine learning algorithms used widely for image recognition tasks. In general, they use a series of convolution filters to capture local features in image-space among all the points in the data. Multiple convolutional layers stacked together allow the network to extract subtle and sophisticated information, such as the non-Gaussian components in the EoR maps. Additionally, convolution layers typically have one or two orders of magnitude fewer parameters compared with dense layers \citep{Li2019}. The CNNs used in this work consist of several convolution layers followed by \edited{a few dense layers, which} are used to analyze the extracted image information and to regress on $\tau$. This is a typical architecture \edited{for predicting} parameters from the images, and is similar to the ones in \citet{Simonyan2014}, \citet{Gillet2018}, and \citet{LaPlante2018}.

Following common practice, we use $3 \times 3$ filters for convolutional layers and pair them with $2 \times 2$ max pooling layers with a stride of 2, which are used to \edited{cut down} the number of parameters \citep{Riesenhuber1999}. Between each convolutional and max pooling layer, batch normalization \citep{Szegedy2015_batchnorm} is adopted to provide regularization. 
\edited{A global average pooling layer is placed to connect convolution layers and the dense layers, which effectively flattens the data output by the convolutional layers.}
For all of the dense layers except for the final one, we use the ``leaky rectified linear unit'' (leaky ReLU) as an activation function \citep{Maas2013_ReLU}. A $20\%$ dropout is adopted to avoid overfitting \citep{dropout}. All the layers are initialized by the ``he normal'' technique \citep{He2014_henormal}. 
\edited{Each network is trained with a batch size of 32 and a learning rate of $10^{-5}$, and trained to minimize the mean squared error (MSE).}
The CNN architectures in this work are implemented using the Keras framework \citep{keras} \edited{with the TensorFlow as the backend \citep{tensorflow}.}

\begin{deluxetable*}{cccccc}
\tablenum{2}\label{table:architecture}
\tablecaption{\edited{Best Architectures of Trained} CNNs}
\tablewidth{0pt}
\tablehead{
\colhead{} & \colhead{CNN~I } & \colhead{CNN~II} & \colhead{CNN~III}}
\startdata
Dataset      & \cmfast\    & \zreion\      & mixed        \\\hline
Convolution   & 64 filters  & 64 filters  & 64 filters  \\
layers        & 128 filters & 128 filters & 128 filters  \\
              & 256 filters & 256 filters & 256 filters  \\
              &             & 512 filters & 512 filters \\ \hline
Dense         & 400 neurons  & 400 neurons  & 400 neurons  \\
layers        & 100 neurons  & 50 neurons  & 100 neurons  \\
              & 20 neurons   & 20 neurons   & 20 neurons    \\ \hline
Output layer & 1 neuron    & 1 neuron    & 1 neuron    \\
\enddata
\tablecomments{We do not show the max pooling layers and dropout layers here. \edited{Full architectures are described} in Section~\ref{sec:Method}.}
\end{deluxetable*}

In principle, $\tau$ as a cosmological parameter depends primarily on the reionization history of the universe, and as such is a model-independent parameter. Nevertheless, the differences between \zreion\ and \cmfast\ may lead to different results when using the different snapshots as input training data.
\edited{In order to understand that impact of underlying models on the machine learning networks, we train three CNNs with \zreion, \cmfast, and mixed snapshots as training datasets to regress the} value of $\tau$\edited{, and label them as CNN~I to III.}
\begin{itemize}
    \item CNN trained and validated exclusively on \cmfast\ data (CNN~I)
    \item CNN trained and validated exclusively on \zreion\ data (CNN~II)
    \item CNN trained and validated on mixed data (including both \zreion\ and \cmfast\ data) (CNN~III)
\end{itemize}

\edited{We make the distinction between ``training data'', which are used during the training process to adjust the weights and biases in the network; ``validation data'', which are used during the training process for monitoring metrics such as the value of the loss function; and ``test data'', which are only used for evaluation after the network has been trained. In the results that follow, we show the performance of the networks on testing data.}

\edited{An optimization process was carried out by a random grid search to decide on the hyperparameters for CNNs~I, II, and III. We varied the number of convolutional layers, the number of filters in the convolutional layers, the number of dense layers, and the number of neurons in these dense layers. We train these networks using a reduced number of epochs to evaluate the validation loss accurately. We identify the ``best'' architectures, which are shown in Table~\ref{table:architecture}, as the ones have the best performance on validation data, i.e., has the smallest validation loss, and reinitialize the network to train with more epochs. Here only the loss on the trained model are considered, i.e., \cmfast\ for CNN I, \zreion\ for CNN II and mixed data for CNN III.}

If the networks are generally independent of these two semi-numeric reionization models, we expect any CNN we mentioned above, no matter which dataset trained on, would work equally well when predicting on test data generated by either \zreion\ or \cmfast. If, instead, the CNN does depend on the model used for generating snapshot data, then there would be a clear difference between the prediction results on these two \edited{methods} for those \edited{networks trained on only one of them}. In that case, the CNNs are implicitly able to distinguish between \zreion\ and \cmfast, even though they are being used to predict on a model-independent parameter $\tau$. Thus, such a result would indicate that these two \edited{methods} have distinct features that do not generalize, and furthermore that CNN-based methods are sensitive to these differences during training.
This result is not ideal because when CNN-based inference methods are to be deployed on actual telescope data, the underlying physics of reionization will certainly be more complicated than either semi-numeric model individually. In order to achieve robust predictions on astrophysical or cosmological parameters, we would like to train models that perform equally well on the two methods. To achieve such even-handed performance, we need to provide the networks with sufficient motivation to work well on both \edited{\zreion\ and \cmfast, i.e., to train} the networks on data from both models. To demonstrate this, we \edited{train the CNN~III with a dataset that contains equal number of \zreion\ and \cmfast\ snapshots, which is supposed to include all the information represented by these two methods.}

\section{Results}
\label{sec:Results}

\begin{figure}[t]
\plotone{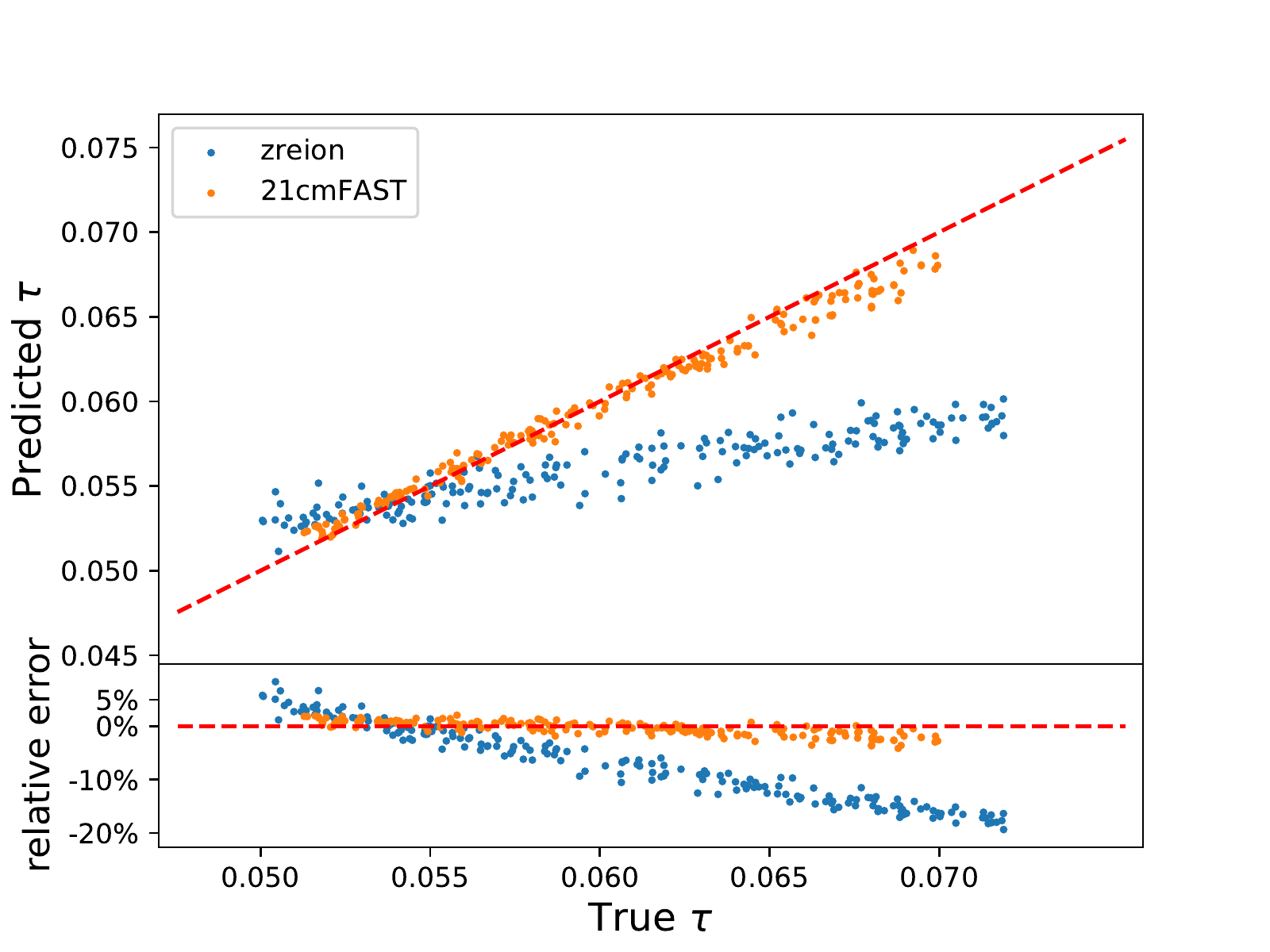}
\caption{Regression results for CNN~I, \edited{which is trained exclusively on \cmfast\ data. 
The orange dots are predictions from an independent \cmfast\ test dataset, and blue dots are from an independent \zreion\ test dataset. The relative error is defined as $|\tau_{\mathrm{true}} - \tau_{\mathrm{predicted}}|/\tau_{\mathrm{true}}$}}
\label{fig:results_21cmfast}
\end{figure}

\begin{figure}[t]
\plotone{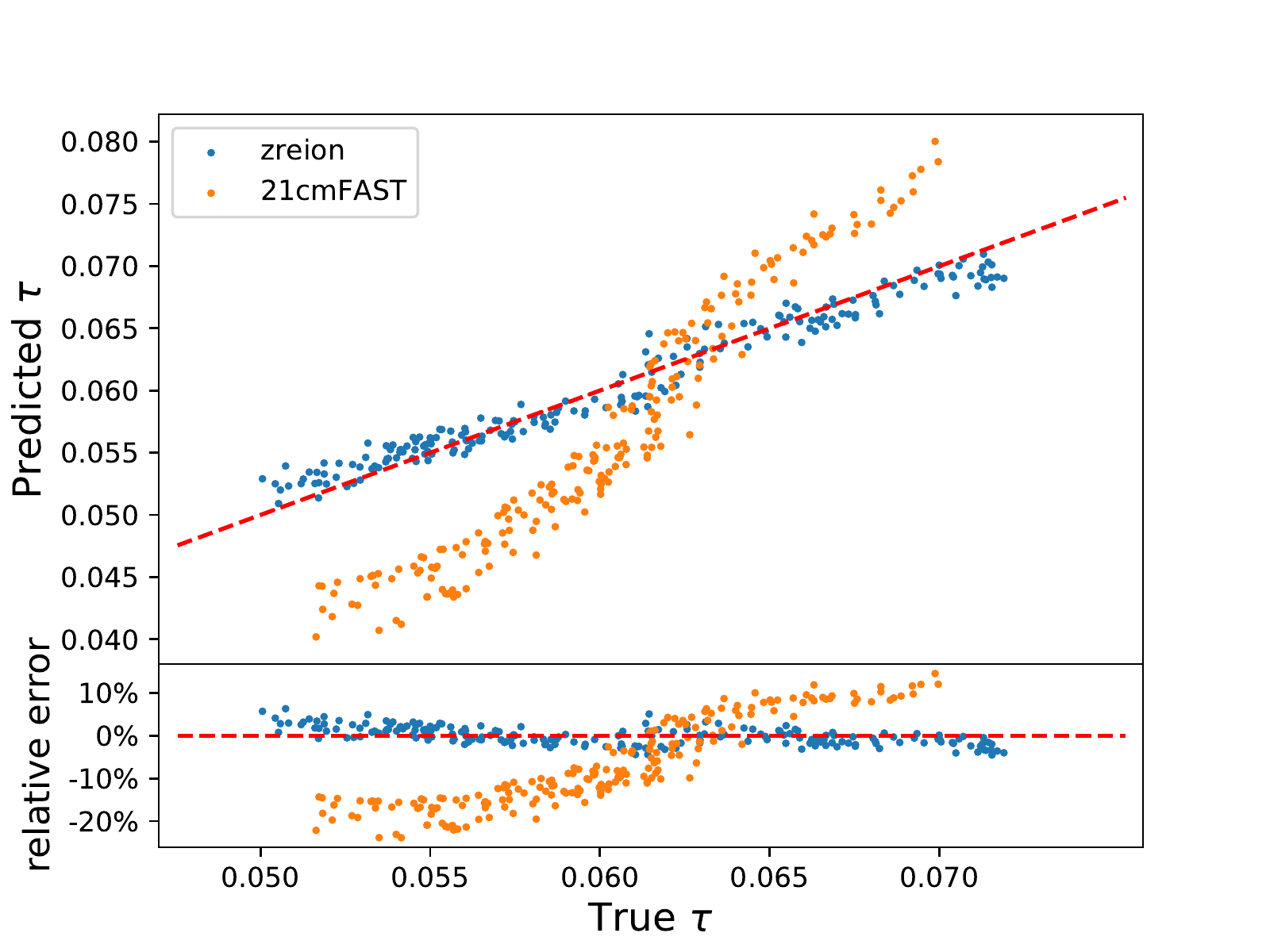}
\caption{Regression results for CNN~II, \edited{which is trained exclusively on \zreion\ data. 
The orange dots are predictions from an independent \cmfast\ test dataset, and blue dots are from an independent \zreion\ test dataset.}}
\label{fig:results_zreion}
\end{figure}

\begin{figure}[t]
\plotone{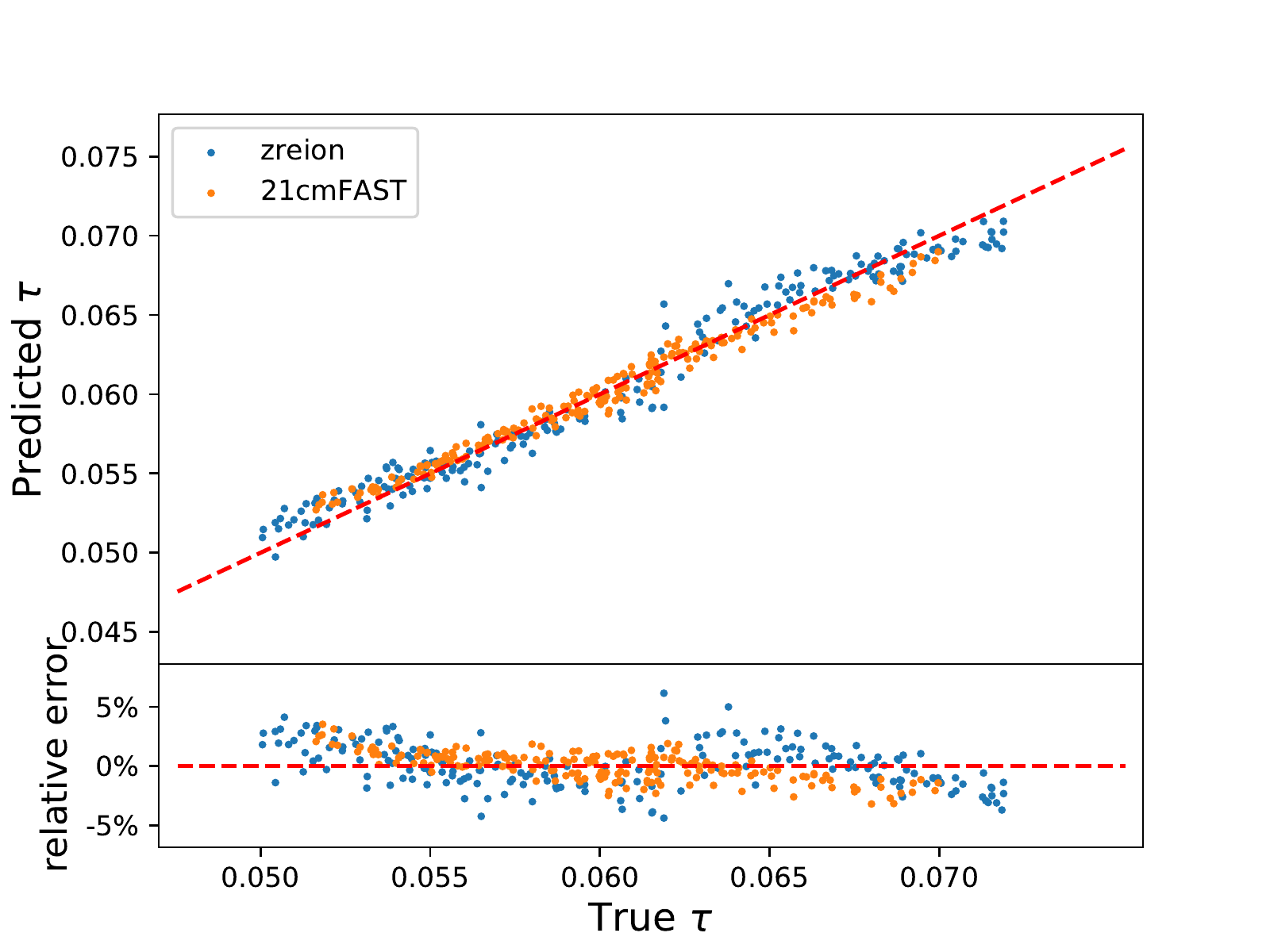}
\caption{Regression results for CNN~III, \edited{which is trained on both \zreion\ and \cmfast\ data. 
The orange dots are predictions from an independent \cmfast\ test dataset, and blue dots are from an independent \zreion\ test dataset.}}\label{fig:mix_results}
\end{figure}

In this section, we present the prediction results of training our CNN networks described in Sec.~\ref{sec:Method}.
We show the architectures that produce the best predictions for the \edited{three} CNNs in Table~\ref{table:architecture}.
Figure~\ref{fig:results_21cmfast} \edited{plots the prediction results made by CNN~I, which is trained on snapshots from \cmfast, on the data from two methods}. These data correspond to a testing dataset generated by each of these \edited{methods}, which were not included in the \edited{training} or validation data.
\edited{The relative error shown in the figures are calculated by $|\tau_{\mathrm{true}} - \tau_{\mathrm{predicted}}|/\tau_{\mathrm{true}}$.}
As can be seen, the network could only produce good predictions on the models they trained on, for which the errors are within 5\%. For the other model, the prediction error ranged from a few percent to over 20\%, with a large bias at high-value end.\edited{Figure~\ref{fig:results_zreion} shows the prediction for CNN~II, which is trained exclusively on \zreion.}
Analogously, this \edited{network} works well on \zreion\ but has large errors on \cmfast. 
\edited{Following this}, the results of CNN~III are shown in 
Figure~\ref{fig:mix_results}, which is trained on a mixed dataset containing snapshots from both \zreion\ and \cmfast. The errors for this network are \edited{all} within 5\%, with no obvious difference between the two \edited{methods}. 

\section{Discussion}
\label{sec:discussion}

\begin{figure}[ht]
\centering
\includegraphics[width=0.95\textwidth]{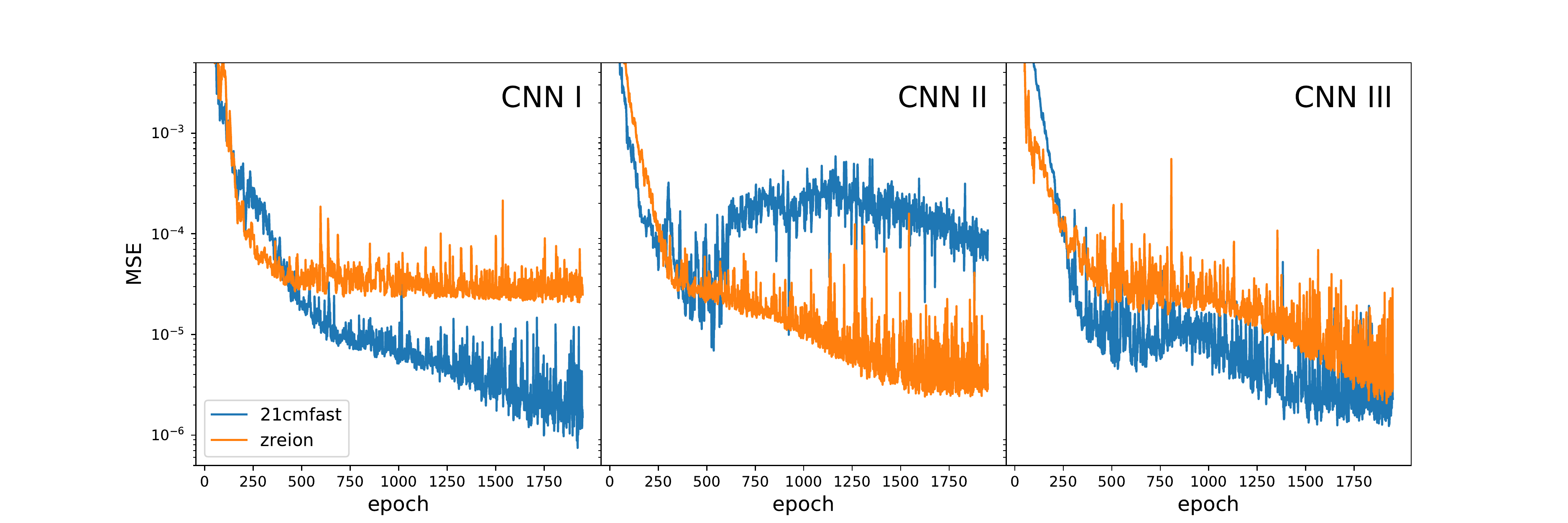}
\caption{The \edited{MSE} loss values \edited{for CNN~I, II, III on \zreion, \cmfast\ data} respectively during the training. For all the subplots, the blue line is the loss \edited{on \cmfast\ and the orange line is the loss on} \zreion. 
}
\label{fig:loss_9CNNs}
\end{figure}

\subsection{CNN's behavior on two models}
\label{section:CNN_behave_compare}

The results shown in Sec.~\ref{sec:Results} indicate that although the snapshots generated by two reionization models are very similar by eye, they are nevertheless \edited{sufficiently different that inhibit CNN from performing equally well on the} both models. 
In this section, we compare \zreion\ and \cmfast\ based on the CNN's behavior by analyzing the loss and comparing \edited{their} architectures in Table~\ref{table:architecture}. \edited{We show the MSE loss on \zreion\ and \cmfast\ respectively during the training for CNNs I, II, and III in Figure~\ref{fig:loss_9CNNs}.
During the early stages, the loss on \zreion\ and \cmfast\ are both decreasing, which implies that the networks are initially identifying features that are common to the two models. 
With additional training epochs, the loss value for \zreion\ and \cmfast\ starts to deviate, which appear for all the three networks. The different behavior on the two methods indicates that at a certain point, the network starts using specific features from the trained model that are not shared by the other for making predictions. If the goal of the CNN training is to produce a generalized network that performs equally well on \cmfast\ and \zreion, it may be possible to use a simpler architecture, or stop training with a relatively small number of epochs. However, based on the fact that the CNN performs poorly on the untrained model and there is a large improvement on the trained one after the point where the loss starts to diverge, we assume the networks could not produce good predictions on neither of two models at that time.}

One \edited{interesting} conclusion is that \cmfast\ is easier for CNNs to capture features and produce good predictions. In Table~\ref{table:architecture}, it can be seen that \edited{the CNN} trained on \cmfast\ \edited{has a} simpler architecture than \edited{that} trained on \zreion. 
\edited{CNN~I, which} has three convolution layers with 64, 128, and 256 filters, \edited{respectively}, could produce similarly good results on the training simulation \edited{model} with CNN~II, which \edited{has} one more layer with 512 filters. \edited{What is more}, Figure~\ref{fig:loss_9CNNs} shows that for \edited{the network} trained on \edited{the} mixed dataset, the loss on \cmfast\ decreases faster and \edited{ends up} smaller than that on \zreion. For CNN~III, it takes the network about 1500 epochs to \edited{give the similar small loss on \zreion\ }which is achieved on \cmfast\ at only 1000 epochs. This result is slightly non-intuitive because based on how convolution layers work, we might assume that CNNs would be more sensitive to features implemented in Fourier space. 

\begin{figure*}
\centering
\includegraphics[width=0.48\textwidth]{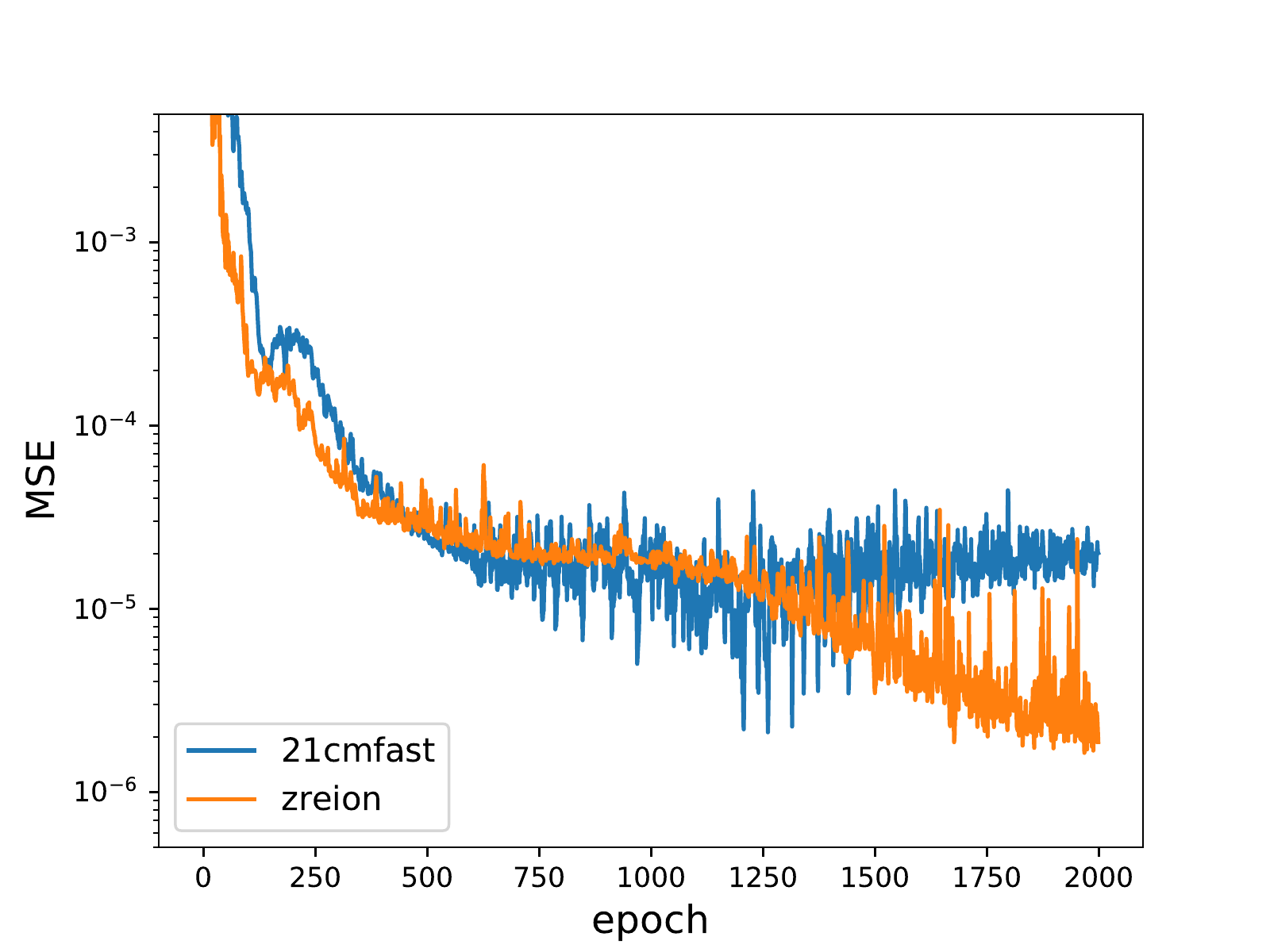}\hfill
\includegraphics[width=0.48\textwidth]{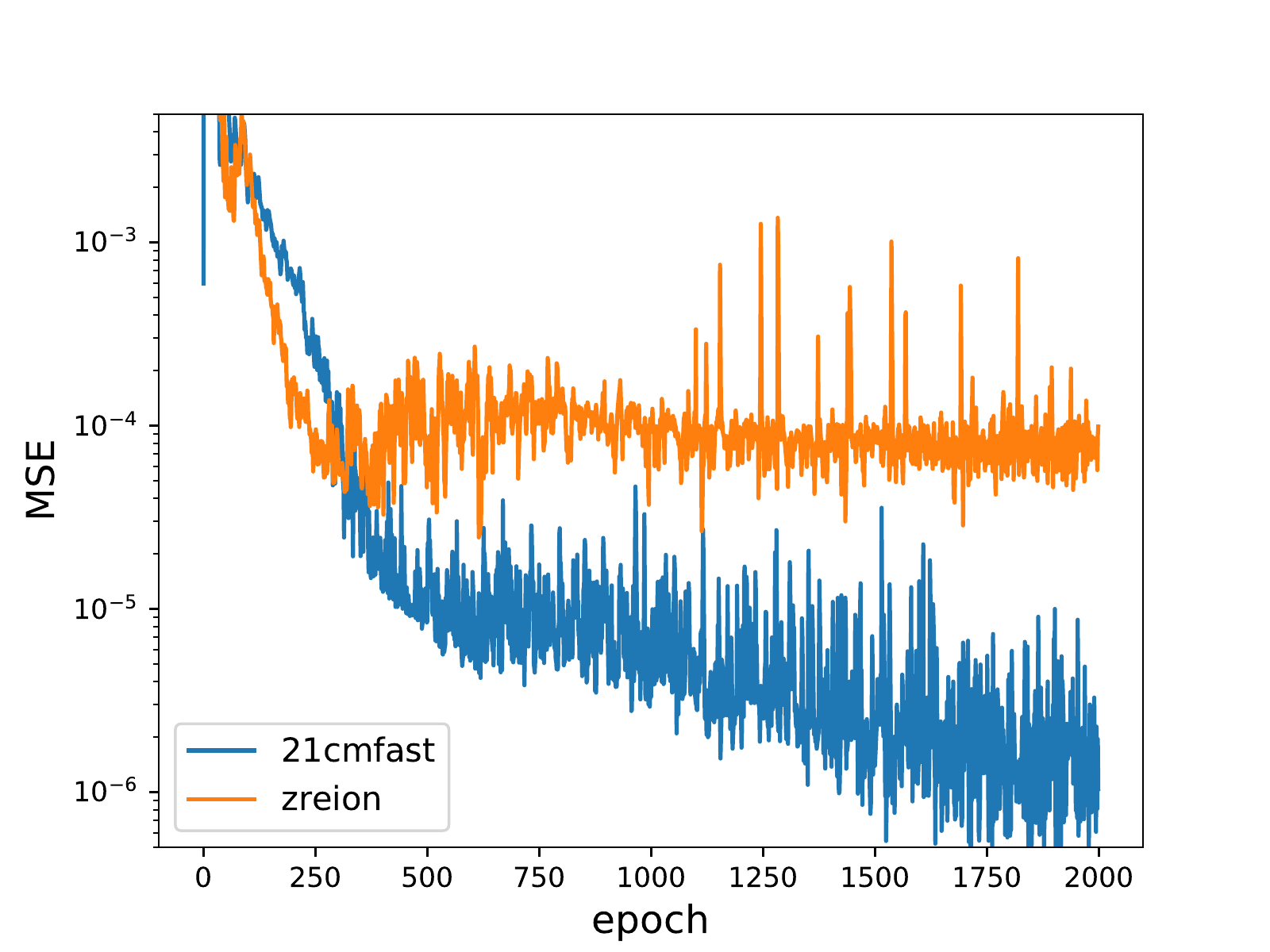}
\caption{
Left: the loss value for the network trained exclusively on \zreion\ with the architecture of CNN~I, which is simpler than that of CNN~II. Right: the loss for the network trained on \cmfast\ with a \edited{more} complex architecture of CNN~II. Their behavior is similar to that shown in Figure~\ref{fig:loss_9CNNs} \edited{except that the overfitting on the left plot is not obvious}, which rule out the possibility of network complexity causing the discrepancy.
\label{fig:archi_check}}
\end{figure*}

We also notice different behavior between CNN~I on \zreion\ and CNN~II on \cmfast. When the network is trained on \cmfast\ (the \edited{left column} in Figure~\ref{fig:loss_9CNNs}), the loss on \zreion\ \edited{stays at the same level} from some point. While for the \edited{network} trained on \zreion\ (the \edited{center column} in Figure~\ref{fig:loss_9CNNs}), \edited{it overfits} on \cmfast\ data. This might be caused by the different architecture complexity since \zreion\ data \edited{need} more convolution layers. \edited{To investigate this, we} use the architecture of CNN~I to train \zreion\ and the architecture of CNN~II to train \cmfast, whose losses are shown in Figure~\ref{fig:archi_check}. The networks cannot learn further \edited{on} \zreion\ even with a more complex architecture, and the overfitting on \cmfast\ still happens in a simple \edited{architecture though it is not as obvious as that shown in Figure~\ref{fig:loss_9CNNs}. These results} rules out the possibility that architecture complexity causes the discrepancy.

\edited{The above} analysis supports the idea that these two models share some common features while have some exclusive characteristics, and the general parts are not enough to allow CNNs to make accurate prediction of $\tau$. \edited{The distinction on loss} might indicate that the relationship between 21\,cm map and resulting value of $\tau$ in \cmfast\ is more straightforward \edited{than that in} \zreion. 
This also \edited{implies} that machine learning might be a potential method to compare different simulation models and draw comparisons between them.

\subsection{Uncertainty Qualification by Monte Carlo Dropout}
\label{label:MC_dropout}

\begin{figure}[t]
  \centering
  \includegraphics[width=0.95\textwidth]{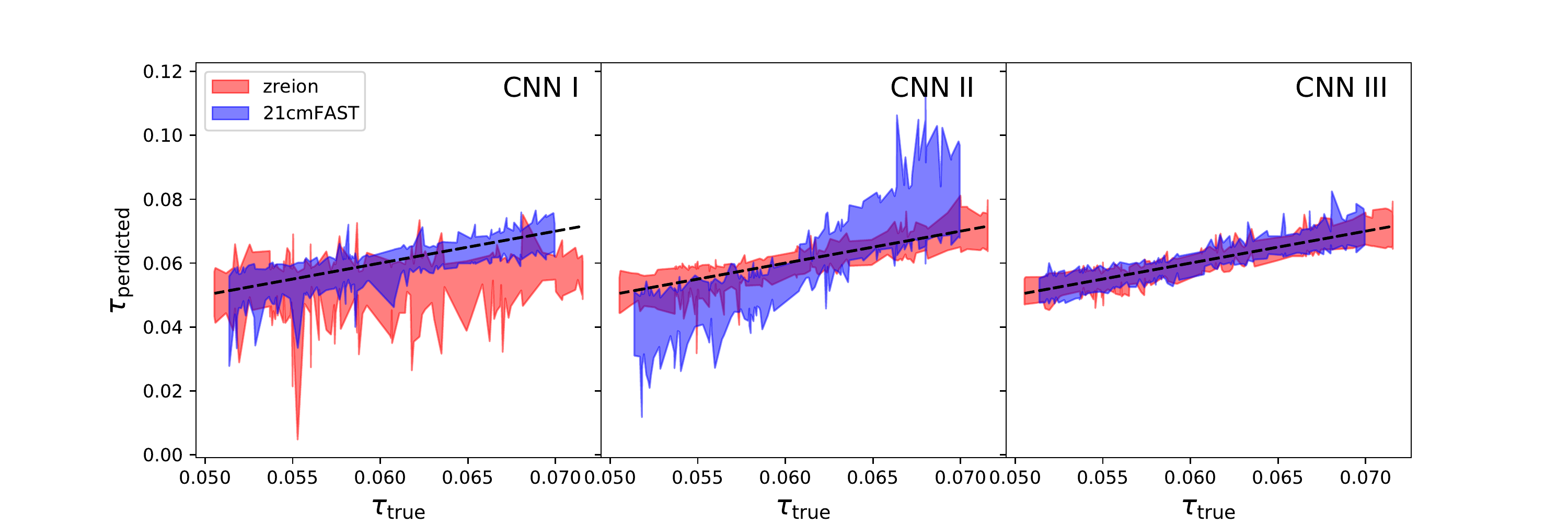}
  \caption{The 1$\sigma$ uncertainty \edited{estimated by Monte Carlo Dropout technique on CNN~I, II, III on} 100 samples for \cmfast\ and \zreion\ respectively. \edited{For each sample 100 predictions are computed.} See Sec.~\ref{label:MC_dropout} for further discussion.}
  \label{fig:MC_dropout}
\end{figure}

In this section, we use Monte Carlo Dropout to estimate the \edited{uncertainties of CNNs' results} on different semi-numeric methods. 
Dropout was first proposed as a method to solve the \edited{overfitting} problem and traditionally is not used during model prediction. However, dropout can also be used as a Bayesian approximation because mathematically neural networks with dropout are equivalent to approximate variational inference in the deep Gaussian process and further \edited{serve} as an effective method to estimate the uncertainty \citep{Brach2020SingleSM, NAIR2020_MCdropout}. Such a technique is called Monte Carlo dropout \citep{MC_dropout_paper}, in which a stochastic process is performed for each input sample because dropout effectively introduces random variables in the networks. As a result, we get an ensemble of predictions for the same input data and the standard deviation of these predictions \edited{is} an estimate of the uncertainty. There are two main sources of uncertainties: aleatoric uncertainties, which represent the inherent property of the dataset, and epistemic uncertainties, which correspond to an inadequate understanding of the underlying models \citep{UQ_review}. The uncertainty caused by the different semi-analytic models falls under epistemic uncertainty, which we seek to quantify here.

\edited{We pick 100 \zreion\ snapshots and 100 \zreion\ snapshots from the test dataset, and use CNN~I, II, III to make 100 predictions on each of the sample, whose standard deviations are plot in Figure \ref{fig:MC_dropout}.}
It can be seen that when the networks trained exclusively on one simulation \edited{method}, the uncertainty on the \edited{untrained model would be larger} across the whole range. And for CNN~III, the one which was trained on a mixed dataset, the uncertainties for \edited{both} \zreion\ and \cmfast\ are small.  
We also show \edited{the averaged standard deviation on the 100 samples $(\sigma_{\cmfast\ } \text{and}\ \sigma_{\zreion})$ for three networks} in Table \ref{table:MC_dropout}. 
We point out that \edited{since the these CNNs networks have different architectures}, the uncertainties within the same \edited{row} should not be compared directly.

\begin{deluxetable*}{cccc}
\tablecaption{\edited{Uncertainty for CNN~I, II, III on \zreion/\cmfast}\label{table:MC_dropout}}
\tablenum{3}
\tablewidth{0pt}
\tablehead{
\colhead{} & \colhead{CNN~I} & \colhead{CNN~II} & \colhead{CNN~III}   
}
\startdata
$\sigma_{\mathrm{\cmfast\ }}$ & 0.0019 & 0.0035 & 0.0014 \\
$\sigma_{\mathrm{\zreion\ }}$   & 0.0026 & 0.0017 & 0.0016 
\enddata
\tablecomments{\edited{These values are got by computing the averaged standard deviation of predictions on 100 \zreion/\cmfast\ snapshots. For each snapshot, 100 predictions are made using Monte Carlo technique.}}
\end{deluxetable*}

\subsection{Larger Convolution Filters}
\label{section:LargeConvFilters}

\begin{figure*}
\centering
\includegraphics[width=0.48\textwidth]{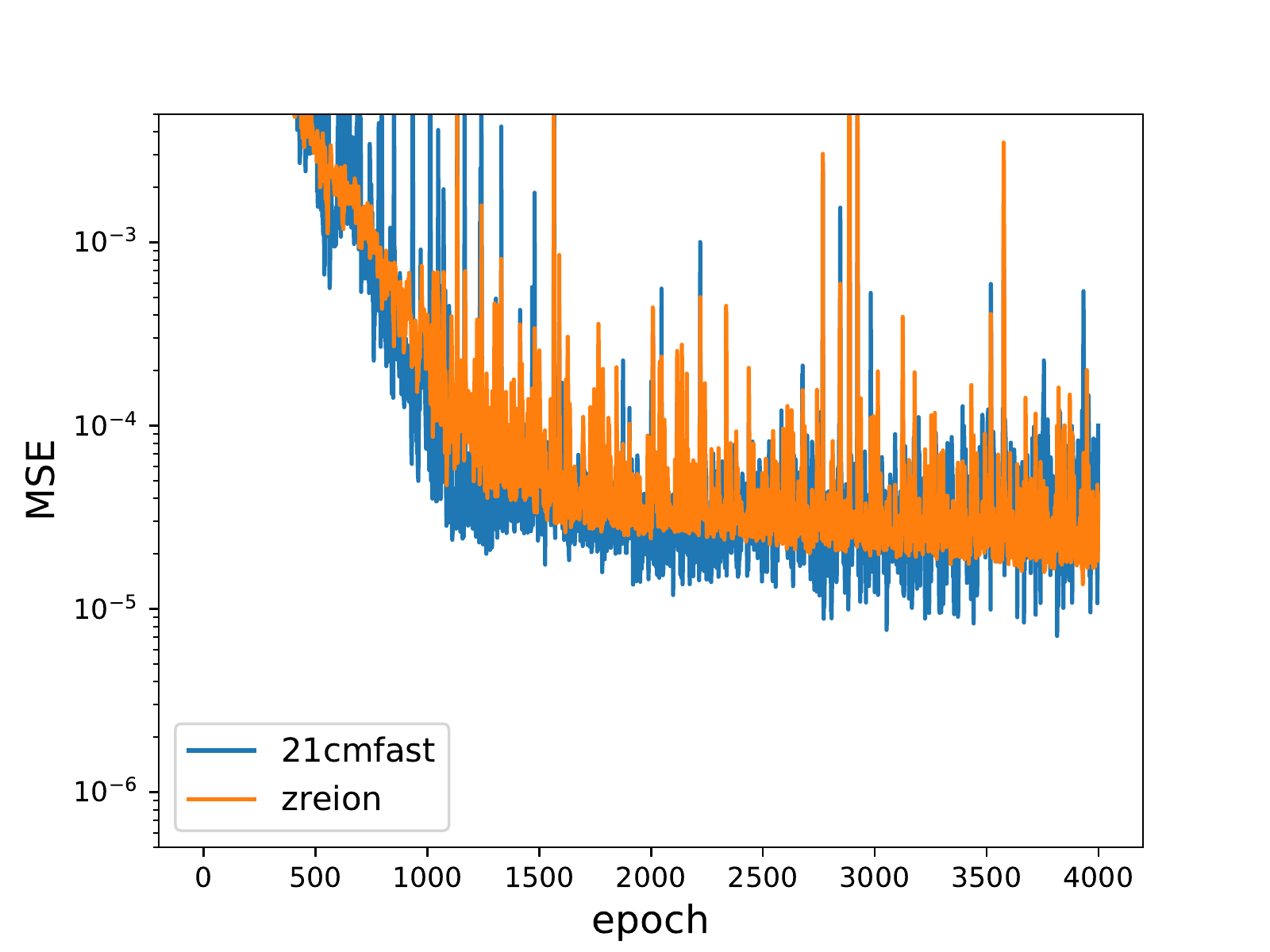}\hfill
\includegraphics[width=0.48\textwidth]{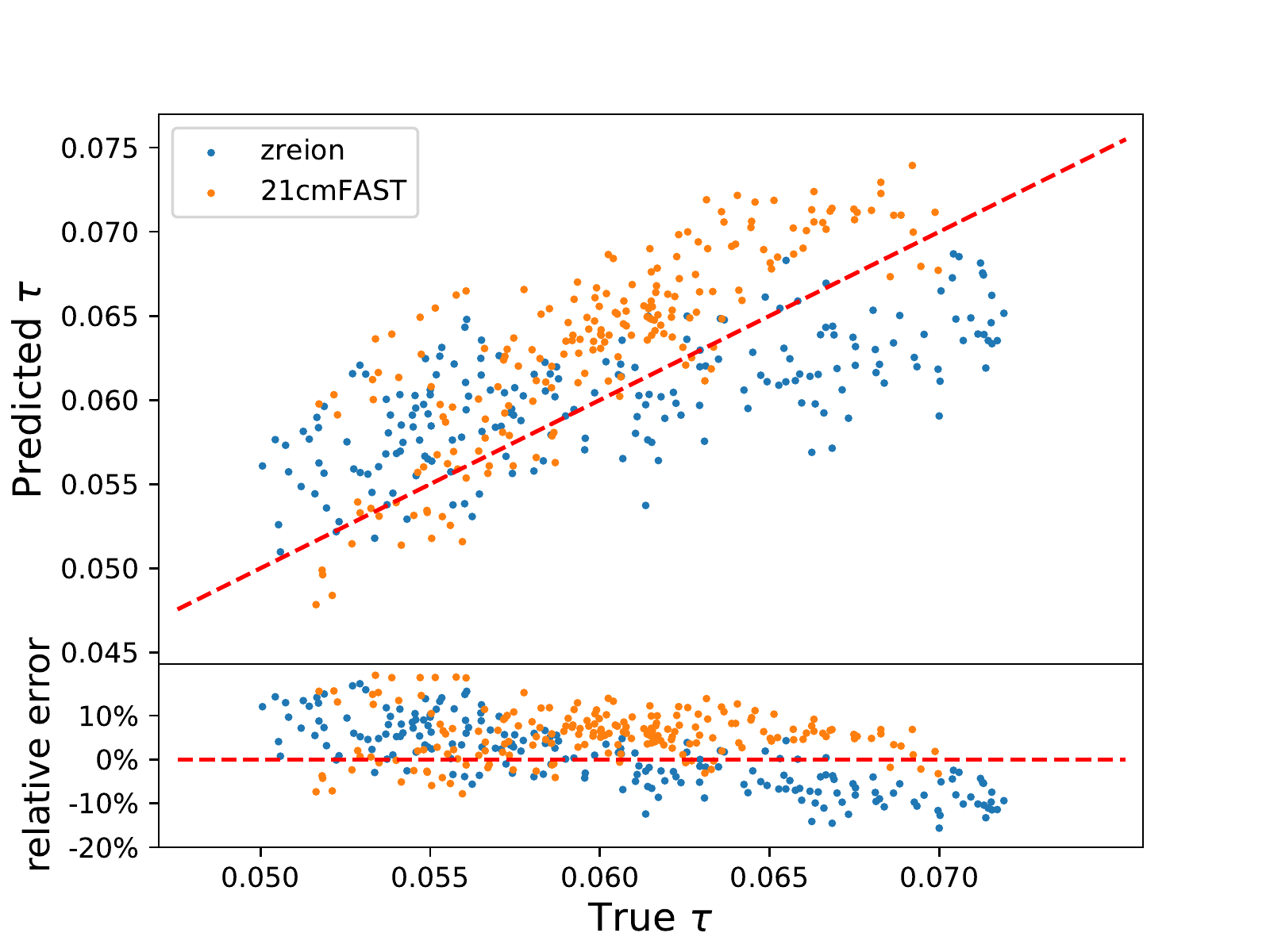}
\caption{
Left: the loss values for \zreion\ and \cmfast\ of a network with $11 \times 11$ convolution filters. The network is trained  exclusively on \zreion\ and has the same architecture of CNN~II shown in the Table~\ref{table:architecture}.
\edited{Right: prediction results for the models using $11 \times 11$ convolution filters. Both simulations have a relatively large scatter in the predicted values.}
\label{fig:large_filters_loss}}
\end{figure*}

For studying EoR, large-scale information in 21\,cm mapping is scientifically interesting. In a CNN \edited{network}, deeper convolution layers\edited{, between which a pooling operation is performed, deal with larger parts} in the original picture, \edited{and in principle,} we could use deep convolution layers to extract large-scale information. \edited{However, small convolution} filters still emphasize more about small-scale \edited{features. As mentioned before, we} use $3\times 3$ filters to train the networks, which \edited{equal} to $11.7 \times 11.7\ h^{-1} \mathrm{Mpc}$ \edited{comoving sizes for the first layer}. Here we try larger filters, $11\times 11$ filters, \edited{which correspond} to $43.0 \times 43.0 \ h^{-1} \mathrm{Mpc}$. Another motivation for using larger filters is that although we have demonstrated CNNs treat \cmfast\ and \zreion\ differently, they might be more similar on larger scales. Training with large filters might give us a network that performs \edited{equally} well on both of the two \edited{methods} even when trained on only one of them. 

One downside with this approach is that a large convolution filter tends to increase the volatility of loss function. As a result, we have to decrease the learning rate, which requires additional training epochs. We use \edited{a} learning rate of $10^{-6}$ with the same architecture of CNN~II shown in Table~\ref{table:architecture}, \edited{and} train exclusively on \zreion\ data. Crucially, we change all the convolution filters from $3\times 3$ to $11 \times 11$.

We plot the loss values for 4000 epochs on \zreion\ and \cmfast\ \edited{snapshots} in Figure~\ref{fig:large_filters_loss}, \edited{which indicates} there is no obvious improvement after 1500 epochs and the loss bottoms out at $10^{-2}$ for both of the two models. \edited{We also show the predictions for the trained networks on snapshots from \zreion\ and \cmfast. As can be seen, the scatter in the predictions in larger than \edited{that} in Figure~\ref{fig:mix_results}.} Meanwhile for CNN~II, whose loss is shown in the \edited{center column of} Figure~\ref{fig:loss_9CNNs}, the \edited{loss value} on \zreion\ keeps going down and ends up less than $10^{-3}$. 
Hence, with larger convolution filters, the gap between \zreion\ and \cmfast\ becomes narrower, while the performance is worse than that of CNNs with smaller convolution filters. When we use $5\times 5$ filters, the predictions are better but the difference \edited{between two} models is still apparent.
This implies that for CNNs, \zreion\ and \cmfast\ do share large-scale features, \edited{which are better captured by larger convolutional filters. Accordingly, larger filters may be used to generate predictions that are more similar between the two methods, at the cost of worse performance on both of them.}

\begin{figure}[ht]
\plotone{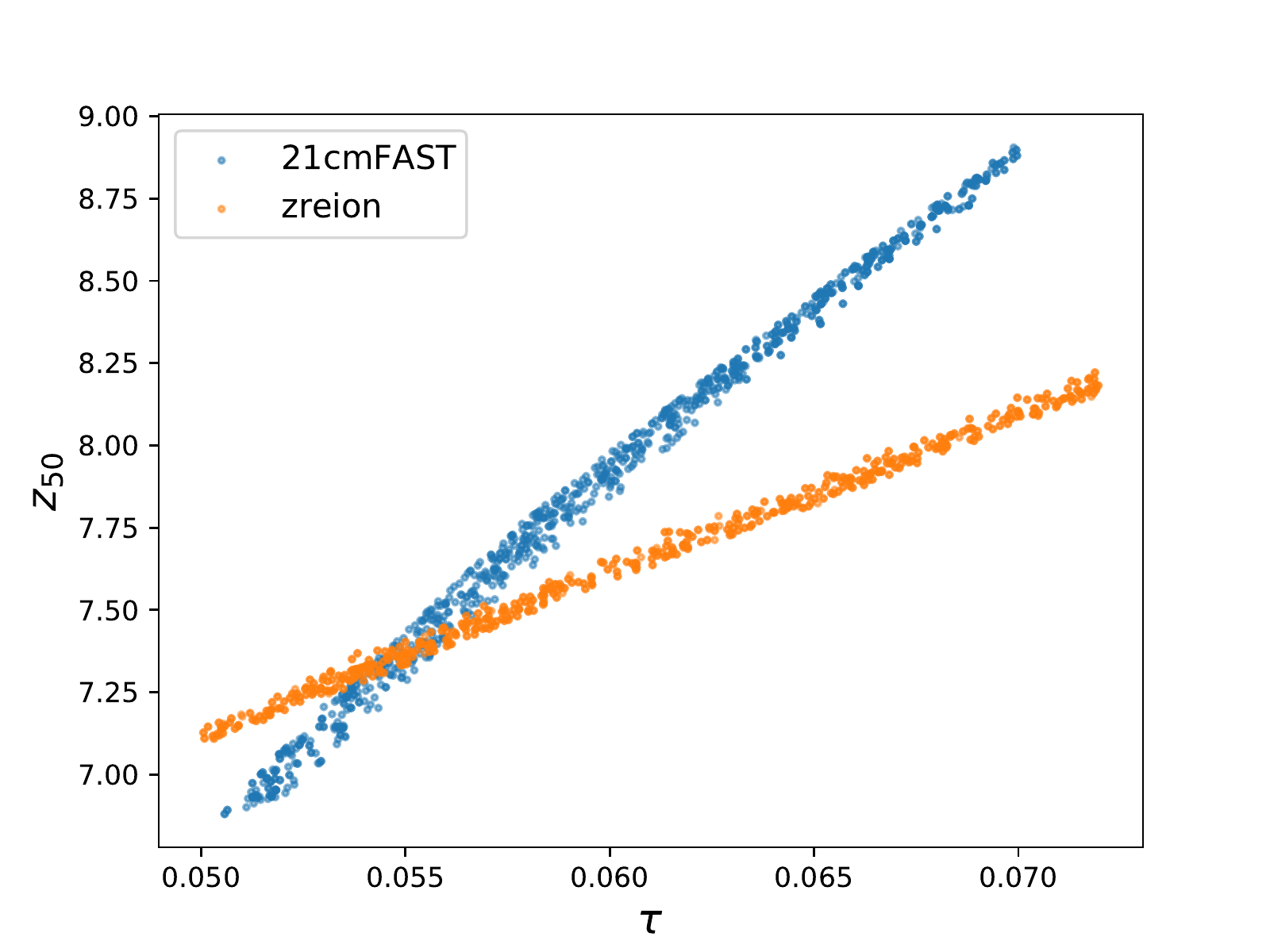}
\caption{\edited{The $z_{50}-\tau$ relations} for \zreion\ and \cmfast, which have obvious divergence.}
\label{fig:z50_tau}
\end{figure}

\subsection{Dependence on redshift}
\label{section:redshift}
The comparison in Sec.~\ref{sec:comparison} shows that the difference between \cmfast\ and \zreion\ is more obvious at low redshift. 
In this section we investigate whether this difference causes \edited{the} discrepancy in \edited{networks'} behavior and if \edited{this redshift range} plays a more important role for CNNs to make decisions.

To visualize the importance of different redshift values, we swap out certain slices in a snapshot and replace them with \edited{the counterpart of the sample from} the other model. 
\edited{To be more specific, we pick several (\zreion, \cmfast) pairs from test datasets and two snapshots in each pair have a similar reionization midpoint $z_{50}$, reionization duration $\Delta z$, and optical depth $\tau$. This selection ensures the different results after swapping mostly come from the substituted slices, rather than distinctions in the ionization history. There are 30 slices in a single snapshot, representing redshift $6<z<15$, and we replace 6 adjacent slices of them each time, which correspond to data around a given redshift.} By feeding these swapped snapshots into CNNs and examining how the prediction \edited{values evolve}, we could have an idea about which redshift range outweighs others for networks to decide optical depth.
In Figure~\ref{fig:z50_tau}, we show that there are different \edited{$z_{50}-\tau$ relations} for \zreion\ and \cmfast. \edited{Accordingly}, the requirement of using similar \edited{$z_{50}$ and $\tau$ means we are only allowed to pick} samples with $\tau \sim 0.055$. 

\begin{figure*}
\centering
\includegraphics[width=0.48\textwidth]{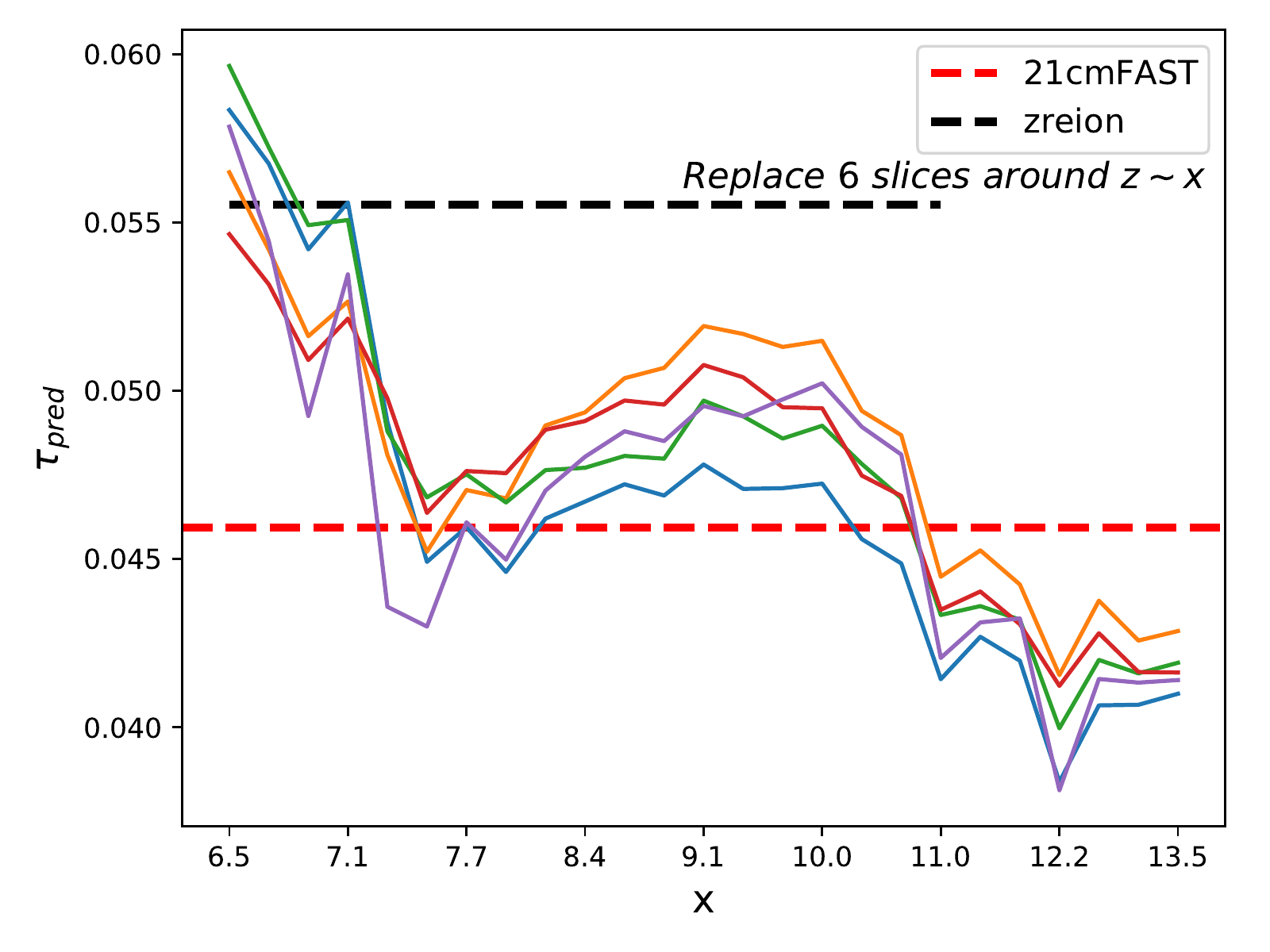}\hfill
\includegraphics[width=0.48\textwidth]{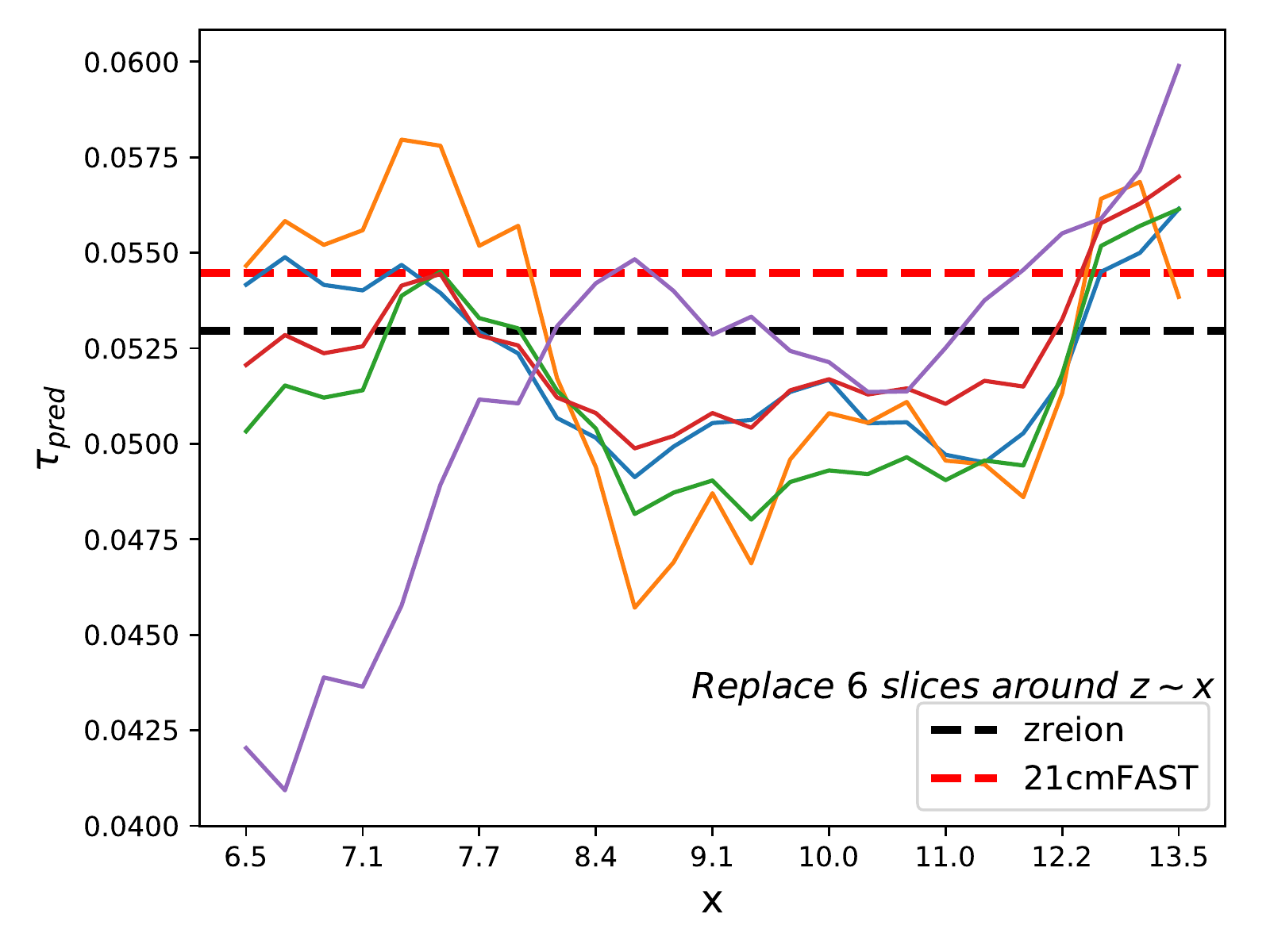}
\caption{
\edited{Left: the evolution of prediction results for 5 swapped \cmfast\ snapshots with 6 adjacent slices from \zreion\ snapshots, which corresponds to data at different redshift values. The predictions are made by CNN~II. Right: the predictions for CNN~I on 5 swapped \zreion\ snapshots with 6 adjacent slices from \cmfast\ data. Each pair of \zreion\ and \cmfast\ samples have similar reionization histories and $\tau$.
The black dashed lines mark the average value of predictions on \edited{the original} \zreion\ samples and the red dashed lines are the average value of predictions on the original \cmfast\ samples.}
}
\label{fig:snapshots_swapping}
\end{figure*}

In the left panel of Figure~\ref{fig:snapshots_swapping}, \edited{we pick 5 pairs of snapshots and use CNN~II to make predictions on these original \zreion\ snapshots and \cmfast\ snapshots, whose averaged values are marked by the black dashed line and red dashed line, respectively.
We note that here the average could serve as a marker of all the initial predictions since all the $\tau_{\mathrm{true}}$ are close to 0.055, which makes the picked \zreion\ or \cmfast\ samples have similar $\tau_{\mathrm{pred}}$. 
Then, we show predicted values for the 5 swapped \cmfast\ snapshots, which have 6 slices from \zreion\ data and are regressed by CNN~II, in which case the black dashed line is our ``target'' result while the red dashed line is the ``original'' result}. 
\edited{The $\tau_{\mathrm{pred}}$ being close to the black line means} the corresponding swapped redshift range is important for accurately predicting on \zreion.
\edited{As can been seen for \zreion, the CNN makes much more use of low-redshift data because} simply replacing the existing \edited{$z<8$ layers in \cmfast\ snapshots} with \zreion\ ones dramatically improves the ability of the CNN to accurately predict the ``correct'' values for $\tau$. 
\edited{In the right panel of Figure~\ref{fig:snapshots_swapping}, we do the same test on 5 swapped \zreion\ snapshots, which have 6 slices from \cmfast\ and are regressed by CNN~I.} From Figure~\ref{fig:results_21cmfast} it is clear that \edited{around} $\tau \sim 0.055$, CNN~I produces almost the same results on two \edited{methods}, which makes the red dashed line and black dashed line are close to each other. Combined with the large scatter, it is difficult to tell which redshift range is crucial for \cmfast. 

\begin{figure*}
\centering
\includegraphics[width=0.48\textwidth]{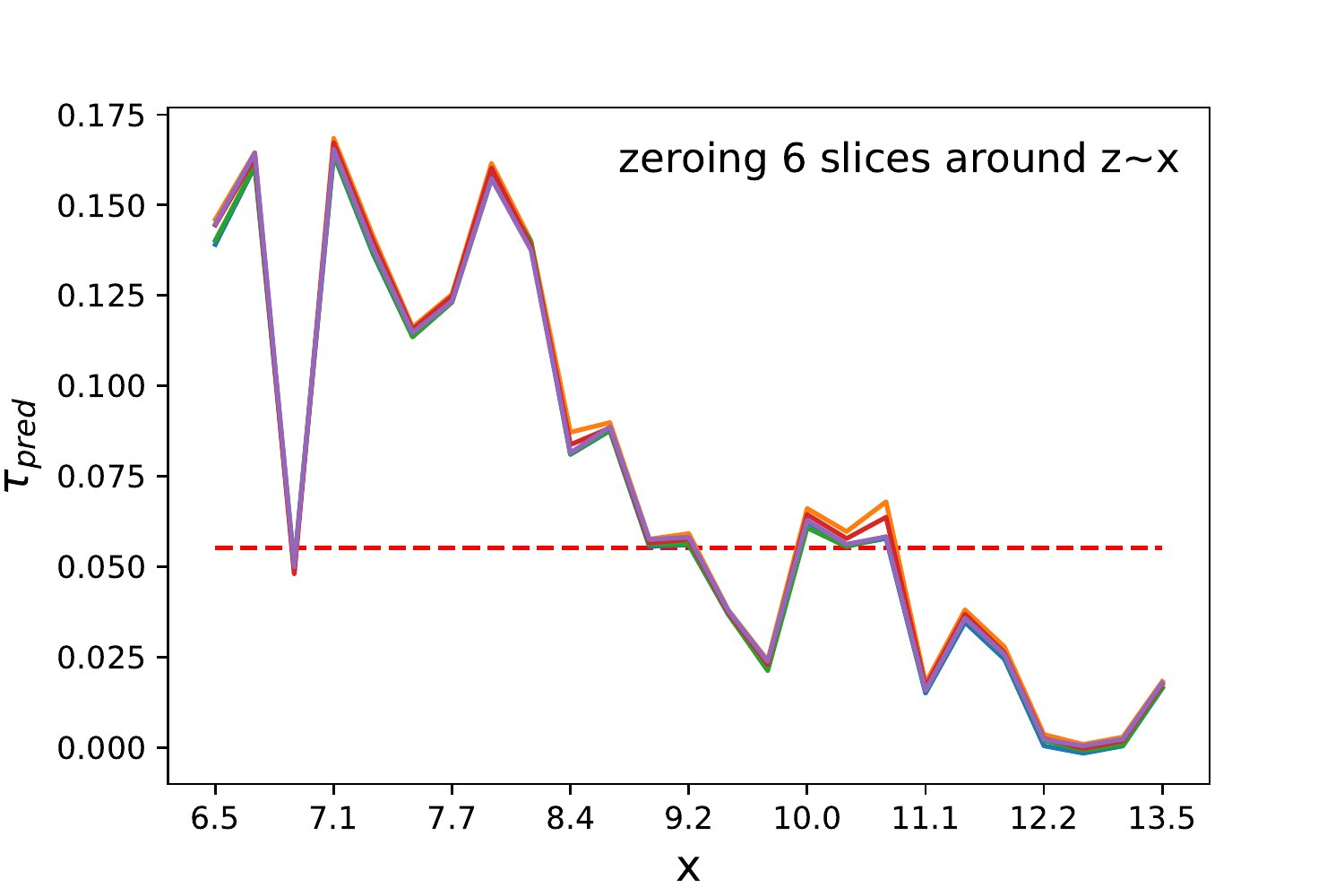}\hfill
\includegraphics[width=0.48\textwidth]{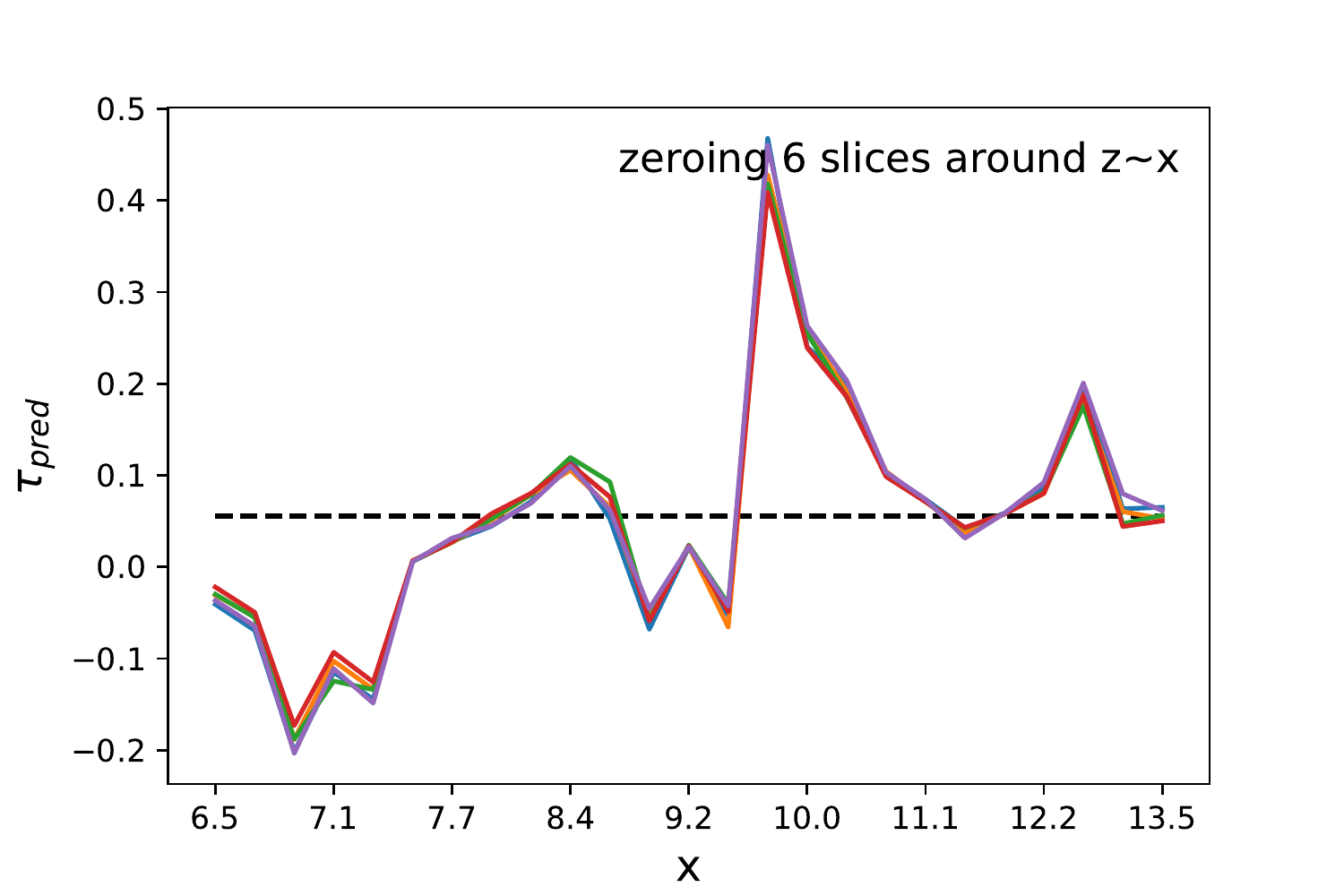}
\caption{\edited{The evolution of prediction results for 5 samples with 6 adjacent zero slices, i.e., removing the information at given redshift values. Left: The predictions made by CNN~I on 5 zeroing \cmfast\ snapshots. The red dashed line marks the average value of predictions on the original \cmfast\ samples. Right: The predictions made by CNN~II on 5 zeroing \zreion\ snapshots. The black dashed line marks the average value of predictions on \edited{the original} \zreion\ samples.
}}
\label{fig:snapshots_zeroing}
\end{figure*}

\edited{As a comparison, we zeroing the same adjacent slices in the snapshots and plot the evolution of predicted values in Figure~\ref{fig:snapshots_zeroing}, where the red/black dashed lines still indicate the average initial predicted values. The deviation from the original predictions is much larger than that shown in Figure~\ref{fig:snapshots_swapping}, which implies the network learn much information from the other model although this is not sufficient to make accurate regression.
When the slices at low redshift are set to zero, the deviation from the `original' predictions is generally larger than that on at high redshift, which again supports the idea that low-redshift data is more crucial for the networks to predict $\tau$.
}

\begin{figure}[ht]
\centering
\includegraphics[width=0.9\textwidth]{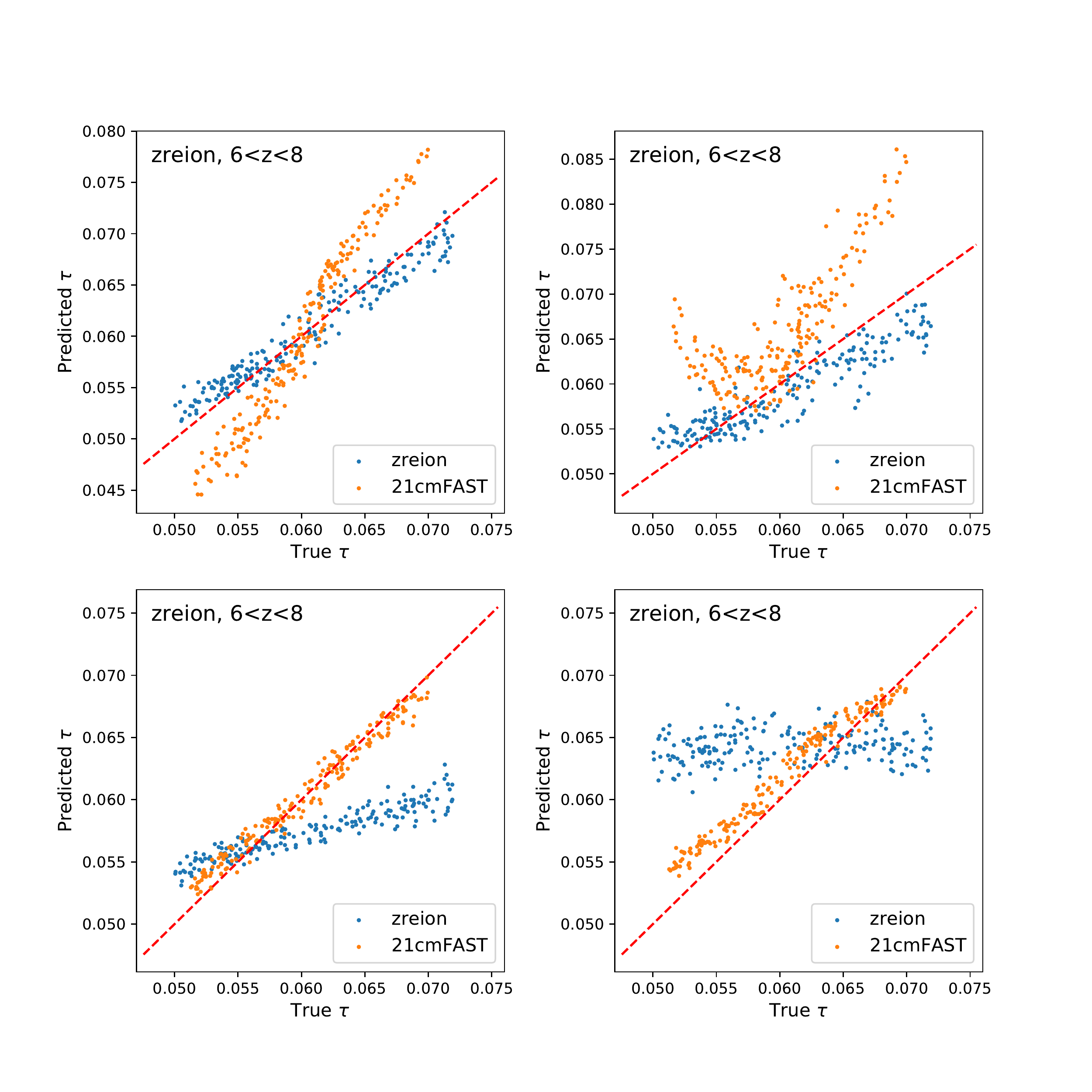}
\caption{The regression results on \zreion\ and \cmfast\ when the training dataset includes information only at low redshift values ($6<z<8$, \edited{11 slices}) or high redshift values ($8<z<15$, \edited{19 slices}). The upper two plots are results for networks trained exclusively on \zreion\ data, \edited{which have the same architecture with CNN~II except for the input data shape}, and the lower two plots are results for networks trained exclusively on \cmfast\ data, \edited{which have the same architecture with CNN~I except for the input size.}}
\label{fig:depen_on_z}
\end{figure}

As another test \edited{of the networks' dependency on each redshift range, we train CNNs exclusively on low redshift part ($z\lesssim8$, 11 slices) or high redshift part ($z >8$, 19 slices) of the snapshots, whose results are plot in Figure~\ref{fig:depen_on_z}.}
Here we use the exact same architecture, learning rate, and number of training epochs for both CNN~I and CNN~II to train \cmfast\ and \zreion\ \edited{data}, so the only thing changed is the input data shape. 
Figure~\ref{fig:depen_on_z} shows that when we only include low-redshift data \edited{in the training}, the networks produce almost the same results \edited{with those in Figure~\ref{fig:results_21cmfast} and \ref{fig:results_zreion}. On the other hand,} when only high-redshift data \edited{are used, the performance gets worse.}
This suggests a similar \edited{conclusion} as for Figure~\ref{fig:snapshots_swapping}, \edited{which is that the information at low redshift plays a more important role for the CNNs to make the regression for both semi-numeric models used.}

\subsection{Visualizing CNN Feature Extraction}
\label{section:filters_vis}

\begin{figure}[ht]
\centering
\includegraphics[width=0.7\textwidth]{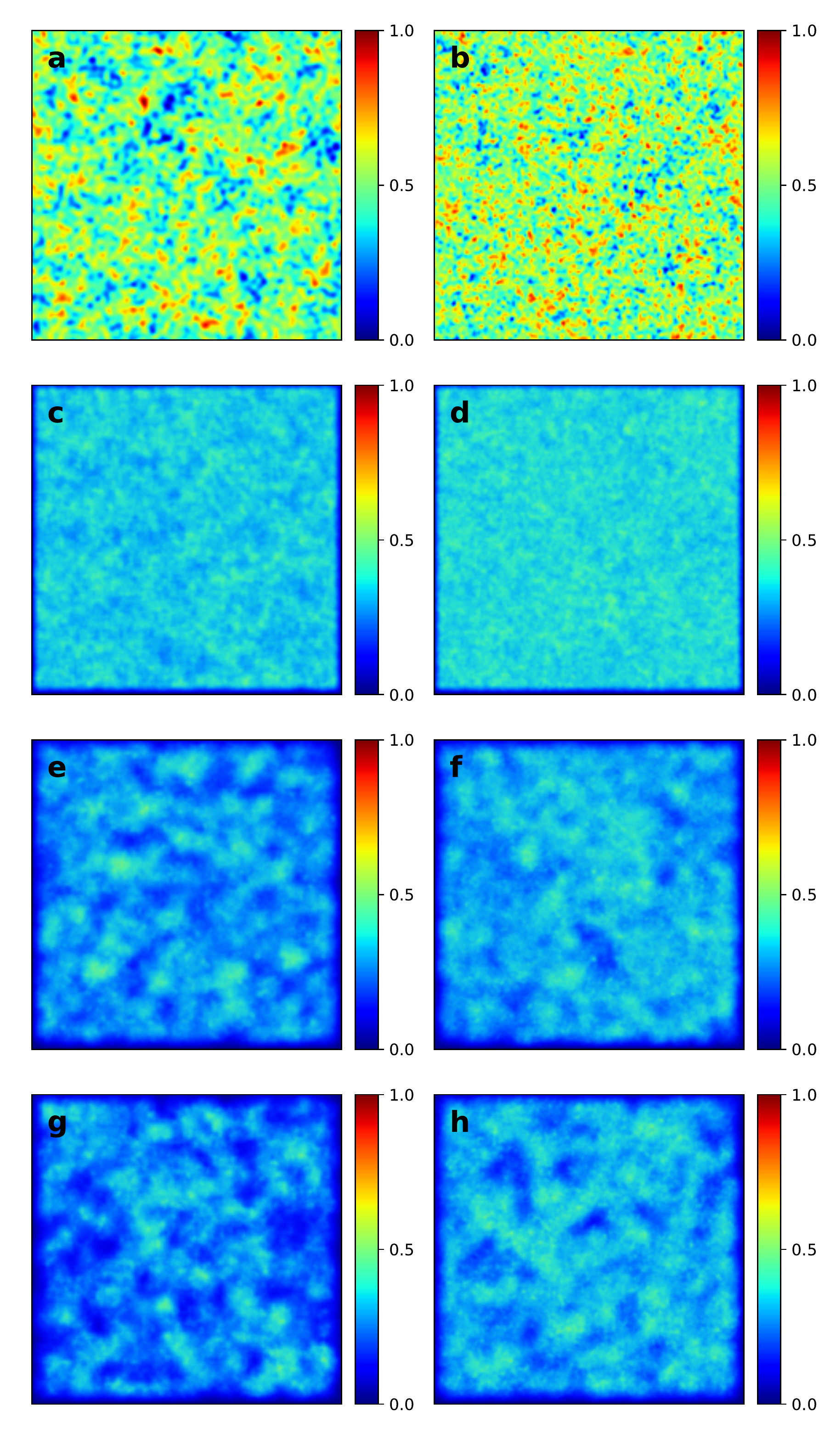}
\caption{The saliency maps of \edited{CNN~I, II, III, and the corresponding snapshots. 
The first row show the central slice ($z\sim 9$) of \cmfast\ snapshot (panel a) and \zreion\ snapshot (panel b) used to generate the saliency maps. The panels (c)/(e)/(g) are the saliency maps on the \cmfast\ data made by CNN~I/II/III, respectively, and the panels (d)/(f)/(g) are the maps on the \zreion\ snapshot made by CNN~I/II/III. The saliency is normalized for each map and the maximum values of the given pixel across all the channels are coded by the color. We apply 10\% Gaussian noise to the original snapshots for 50 samples and average the gradients to obtain these saliency maps.}}
\label{fig:SaliencyMap}
\end{figure}

Understanding the CNN's interpretation of input data might provide some physical insight into the question itself. There are several methods that have been developed for this purpose, one of which is known as activation maximization \citep{mahendran_vedaldi2015,LaPlante2018,billings2021}. This technique transforms a random input image into one that maximizes the response of the chosen neuron through gradient ascent. The output result indicates which features are important to the network in order to discriminate between different parameter values. 
In this work, we use another way to visualize feature extraction called the saliency map \citep{Tong2010_SaliencyMap, Jiang2013_SaliencyMap}, \edited{in which} we calculate how much the prediction value for a trained network varies when each pixel in \edited{a given} image is perturbed. In other words, this map visualizes the gradient of the prediction with respect \edited{to the} changes in the pixel of the input data. 
Similar to the activation maximization technique, saliency maps were first proposed to analyze classification in CNNs, showing the key features a network uses to identify a specific class. In our regression case, the map indicates the differences between the snapshots with large and small $\tau$ values captured by the CNNs. 

\edited{We calculated the saliency maps for a \zreion\ and \cmfast\ snapshot using CNN~I, II and III, which are shown in Figure~\ref{fig:SaliencyMap}. Panel (a) is the central slice ($z\sim 9$) of the \cmfast\ snapshot  used to generated the maps in (c), (e), (g), and panel (b) is the central slice of the \zreion\ snapshot for maps (d), (f), (h). The ($\tau_{\mathrm{true}}$, $\tau_{\mathrm{CNNI}}$, $\tau_{\mathrm{CNNII}}$, $\tau_{\mathrm{CNNIII}}$) for the \cmfast\ and \zreion\ samples are (0.067, 0.065, 0.074, 0.066) and (0.053, 0.054, 0.053, 0.052), respectively.
The second/third/forth rows are saliency maps for CNN~I/II/III, respectively. To improve the maps' behavior, we use the technique of `SmoothGrad' \citep{smilkov2017smoothgrad} to reduce the visual noise. We add Gaussian noise $\mathcal{N}(0, \sigma^{2})$ on the original snapshots and calculate the average gradients of 50 samples, where $\sigma = 0.1(x_{\mathrm{max}} - x_{\mathrm{min}})$ and the $x$ is the pixel values in the input data. The saliency map is normalized to range from 0 to 1 and the color is coded by the maximum values of the given pixel across all the channels}.

The gradient of CNN~I (\edited{panels (c) and (d)}) is generally larger than \edited{that of} others, and the distribution of power is more isotropic and accentuates more small-scale features. 
The maps for CNN~II (\edited{panels (e) and (f)}) have large-scale patchy structure, and the contrasts of large and small scales are more obvious, \edited{which} also appears in the maps for CNN~III (\edited{panels (g) and (h)}). It is clear to see that the same snapshot produces distinct saliency maps for each network, which implies the individual networks use different sets of spatial features to make their determination. Interestingly, there is no obvious correlation between the \edited{four} plots within a column: the voids and patches that appear in \edited{snapshots} do not coincide with those that appear in \edited{saliency maps, ruling} out the possibility that the structures in \edited{the} maps correspond to the positions of ionized bubbles in snapshots. 

Furthermore, the saliency map for a given network identifies different features for \zreion\ and \cmfast. \edited{Since the saliency map works better for samples with accurate predictions, we concentrate our analysis on panels (g) and (h), which are the saliency maps for CNN~III on the two methods.}
Although the sizes of features are similar \edited{in (g) and (h)}, the contrasts between patches and voids are different, \edited{which indicate that even for the CNN} trained on a mixed dataset, it can still \edited{implicitly distinguish} \cmfast\ and \zreion. One possible interpretation of this result is that the good performance of CNN~III is due to learning the relevant details for both simulations, rather than the shared or common features. If this is the case, then making a more general network might be possible by using adversarial techniques to train a network that is agnostic to the particular details. \edited{We} save such investigations for future work.

\section{Conclusion}
\label{sec:conclusion}

In this work, we investigate the extent to which training CNNs for reionization applications depends on the underlying semi-numeric model used to generate input snapshots. We \edited{use two semi-numeric methods:} \cmfast\ and \zreion, and \edited{find} that CNNs \edited{are} indeed sensitive to the input \edited{model}. In general, a network trained on data from one model does not \edited{make} accurate predictions for data generated by the other one, even though the parameter $\tau$ is in principle model independent. We find that using a training dataset that contains snapshots from both simulation \edited{methods} leads to satisfactory results when training, though the CNN still implicitly can tell the difference between the snapshots \edited{from the two models}, as suggested by the saliency maps used to understand the networks' behavior.

For future analysis, we can make use of adversarial techniques for reducing the dependence of the trained networks on features unique to a single simulation. For example, by using a discriminator network which attempts to distinguish whether particular input data was generated by \cmfast\ or \zreion, it may be possible to train a network that is not overly reliant on particular details for one simulation over another. Such an approach may be a way to build a ``general'' network that does not over-learn the details of any one network and does not require an arbitrarily large set of semi-numeric models to yield robust estimates. \edited{Such a general} network is essential for making predictions on model-independent parameters, and can help ensure that results are not biased.

Although the particular applications in this work \edited{use} CNNs and map-space data to investigate the dependence on different semi-numeric models, the \edited{distinctions} are also be present in Fourier space, and hence appear in the power spectrum as well (see Figure~\ref{fig:power_spectrum}). When interpreting the results of upcoming experiments like HERA or the SKA, it is important to understand the impact that the particular simulation framework used to model the signal affects the systematic errors associated with the inferred parameters. As such, using a variety of simulation frameworks, like the analysis performed in \citet{h1c_theory}, is essential for building confidence in the ultimate results.

\begin{acknowledgments}
We thank James Aguirre, Adrian Liu, Jordan Mirocha, Julian Muñoz, Michelle Ntampaka, and Jonathan Pober for insightful comments and useful suggestions on this work, and we thank Emma Bloomfield for invaluable feedback on the manuscript. P.L. acknowledges support from the Berkeley Center for Cosmological Physics. This material is based upon work supported by the National Science Foundation under grant Nos. 1636646 and 1836019 and institutional support from the HERA collaboration partners.  This research is funded in part by the Gordon and Betty Moore Foundation through grant GBMF5212 to the Massachusetts Institute of Technology. This work used the Extreme Science and Engineering Discovery Environment (XSEDE), which is supported by National Science Foundation grant No. ACI-1548562 \citep{xsede2014}. Specifically, it used the Bridges-2 system, which is supported by NSF award No. ACI-1445606, at the Pittsburgh Supercomputing Center (PSC, \citealt{bridges2015}).
\end{acknowledgments}

\appendix

\section{Dependence on Cosmological Parameters}
\label{section:cosmology}

\begin{figure}[ht]
\centering
\includegraphics[width=0.9\textwidth]{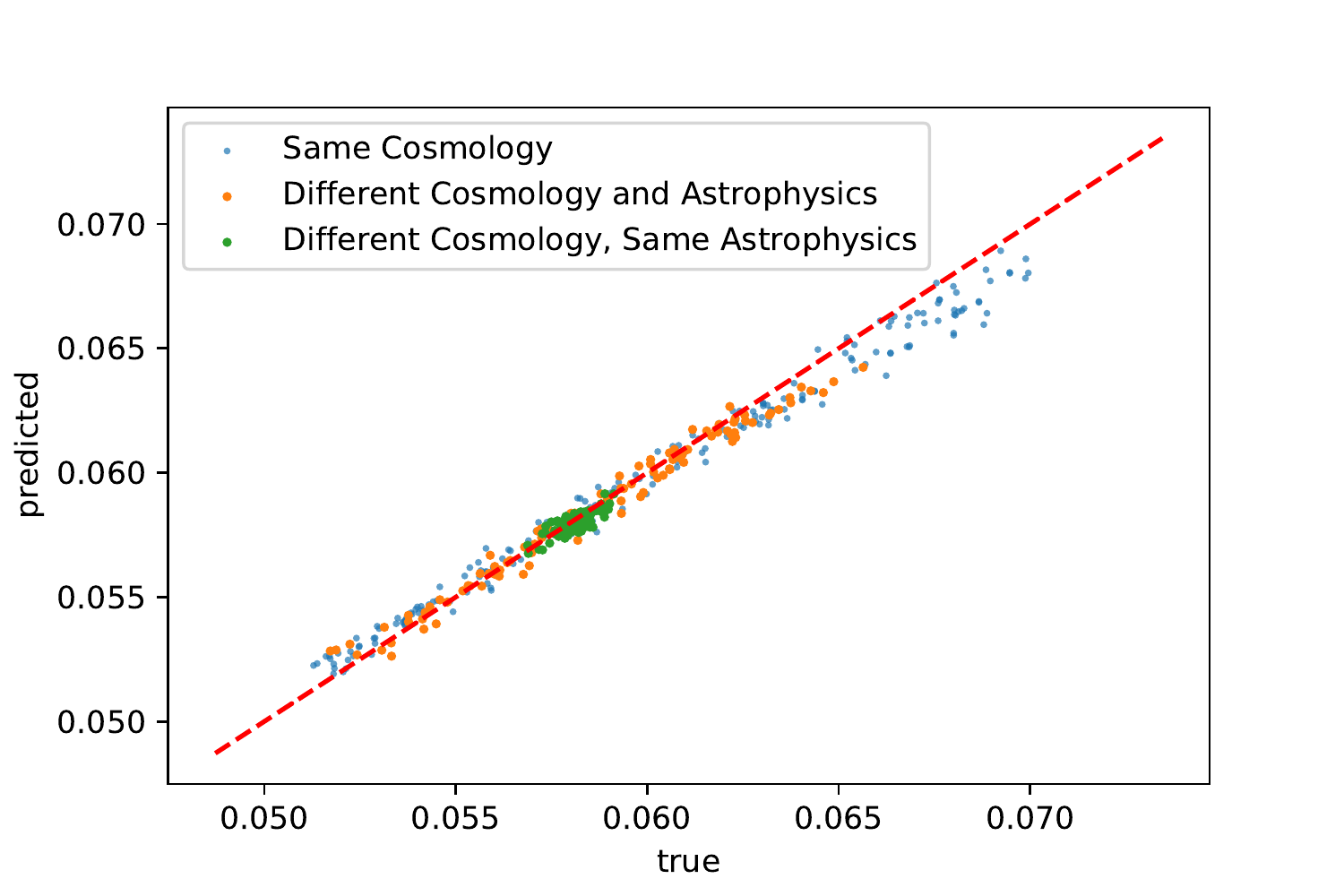}
\caption{\edited{The predictions of CNN~I on \cmfast\ snapshots in the test dataset (blue dots), i.e., with the fiducial cosmology model and different astrophysical parameters, the \cmfast\ snapshots with the different cosmological and astrophysical parameters (orange dots), and the \cmfast\ snapshots with the different cosmological while the same astrophysical parameters (green dots).
}} \label{fig:cosmo_21cmfast}
\end{figure}

Because there are finite errors associated with the underlying cosmological parameters, we investigate the extent to which our CNNs depend on these errors. To do this, we generate 200 \cmfast\ input snapshots with different cosmological parameters \edited{($H_{0}, \Omega_{m}, \Omega_{b}, \sigma_{8}$, $n_{s}$), whose ranges are listed in Table~\ref{table:cosmoparams}.}
We sample each parameter according to Gaussian distributions, whose fiducial set is the is the one used in the training dataset and variances are consistent with the errors in \citet{Planck2015}. \edited{100 snapshots have the same astrophysical parameters while the other 100 snapshots have different astrophysical parameters.
We use CNN~I, which is} trained on our fiducial cosmology \edited{model,} to predict optical depth on them and 
present our results in Figure~\ref{fig:cosmo_21cmfast}.

\begin{deluxetable*}{ccl}
\tablenum{4}\label{table:cosmoparams}
\tablecaption{The \edited{Varied Cosmological Parameters} }
\tablewidth{0pt}
\tablehead{
\colhead{} & \colhead{\edited{Fiducial Value}} & \colhead{$\sigma$}
}
\startdata
$h$          & 0.0674        & 0.0054    \\
$\Omega_{b}$ & 0.0493        & 0.0001   \\
$\Omega_{m}$ & 0.3153        & 0.0073    \\
$\sigma_{8}$ & 0.8111        & 0.0060    \\
$n_{s}$      & 0.9649        & 0.0042   \\
\enddata
\end{deluxetable*}

As can be seen, the optical depth is mainly determined by astrophysical parameters, \edited{since it is primarily determined} on the ionization fraction as a function of redshift $x_i(z)$, and only weakly depends on the cosmological parameters. Figure~\ref{fig:cosmo_21cmfast} shows that the snapshots with the same astrophysical parameters but varied cosmological parameters (green points) fall into a relatively narrow range of $\tau$ values, while those with different astrophysics span a much broader range. We also see that CNNs still produce good results even when the underlying cosmological parameters differ. The \edited{predictions} show slightly more scatter than \edited{that on snapshots with} the same fiducial cosmology, but are still close to the true values. Thus, the additional error associated with changing cosmological parameters is sub-dominant to the error associated with changing the astrophysical parameters. Therefore, for the real data, whose cosmology model might have some deviation from the one used in our training dataset, our CNNs should still perform well.

\bibliographystyle{aasjournal}
\bibliography{ref}

\end{document}